\documentclass{article}
\usepackage{graphicx}

\newcommand{\twoldots}{\mathinner{\ldotp\ldotp}}

\begin{document}

\title{Local Heuristic for the Refinement of Multi-Path Routing in Wireless Mesh
Networks}

\author{Fabio~R.~J.~Vieira$^{1,2,}$\thanks{Corresponding author (fjimenez@cos.ufrj.br).}\\
Jos\'e~F.~de~Rezende$^3$\\
Valmir~C.~Barbosa$^1$\\
Serge~Fdida$^2$\\
\\
$^1$Programa de Engenharia de Sistemas e Computa\c c\~ao, COPPE\\
Universidade Federal do Rio de Janeiro\\
Caixa Postal 68511, 21941-972 Rio de Janeiro - RJ, Brazil\\
$^2$Laboratoire d'Informatique de Paris 6\\
4, Place Jussieu, 75252 Paris Cedex 05, France\\
$^3$Programa de Engenharia El\'etrica, COPPE\\
Universidade Federal do Rio de Janeiro\\
Caixa Postal 68504, 21941-972 Rio de Janeiro - RJ, Brazil}

\date{}

\maketitle

\begin{abstract}
We consider wireless mesh networks and the problem of routing end-to-end traffic
over multiple paths for the same origin-destination pair with minimal
interference. We introduce a heuristic for path determination with two
distinguishing characteristics. First, it works by refining an extant set of
paths, determined previously by a single- or multi-path routing algorithm.
Second, it is totally local, in the sense that it can be run by each of the
origins on information that is available no farther than the node's immediate
neighborhood. We have conducted extensive computational experiments with the new
heuristic, using AODV and OLSR, as well as their multi-path variants, as
underlying routing methods. For two different CSMA settings (as implemented by
802.11) and one TDMA setting running a path-oriented link scheduling algorithm,
we have demonstrated that the new heuristic is capable of improving the average
throughput network-wide. When working from the paths generated by the multi-path
routing algorithms, the heuristic is also capable to provide a more evenly
distributed traffic pattern.

\bigskip
\noindent
\textbf{Keywords:} Wireless mesh networks, Multi-path routing, Path coupling,
Disjoint paths, Mutual interference.
\end{abstract}

\newpage
\section{Introduction}\label{sec:intro}

Wireless mesh networks (WMNs) have lately been recognized as having great
potential to provide the necessary networking infrastructure for communities and
companies, as well as to help address the problem of providing last-mile
connections to the Internet \cite{Nandiraju2007,Siekkinen2007}. However, mutual
radio interference among the network's nodes can easily reduce the throughput as
network density grows above a certain threshold \cite{Balachandran2005} and
therefore compromise the entire endeavor. Such interference is caused by the
attempted concomitant communication among nodes of the same network and
constitutes the most common cause of the network's throughput's falling short of
being satisfactory (hardly reaching a fraction of that of a wired network
\cite{Gupta2000}). A promising approach to tackle the reduction of mutual
interference seems to be to combine routing algorithms with some interference
avoidance approach, such as power control, link scheduling, or the use of
multi-channel radios \cite{Akyildiz2005}. In fact, this type of network
interference problem has been addressed by a considerable number of different
strategies to be found in the literature
\cite{Ying2000,Cruz2003,Abolhasan2004,Sheriff2006,Campista2008,Wang2008,Srikanth2010,Augusto2011}.

An alternative approach that presents itself naturally is the use of multi-path
routing to distribute traffic among multiple paths sharing the same origin and
the same destination, since in principle it can help with both path recovery and
load balancing better than the use of single-path strategies. It may, in
addition, lead to better throughput values over the entire network
\cite{Salma2006,Augusto2010}. But while these benefits accrue only insofar as
they relate to how the multiple paths interfere with one another
\cite{Tsai2006,Tarique2009}, unfortunately this aspect of the problem is not
commonly addressed by multi-path strategies. What happens as a consequence is
that, though promising by virtue of adopting multiple paths to accommodate the
same end-to-end traffic, in general such strategies fail to perform as desired
because they do not tackle the interference problem during path discovery. The
single noteworthy exception here seems to be the algorithm reported in
\cite{Waharte2006}, but it uses geographic information (like localization aided
by GPS) to find paths with sufficient spatial separation so as not to interfere
with one another. In our view this weakens the approach somewhat, since such
type of information may not always be available \cite{Demirkol2006}. Moreover,
the corresponding algorithm relies on the solution of an NP-hard problem on an
input that has the size of the network \cite{Waharte2008}, so the solution may
be unattainable in practice.

Here we propose a different approach to alleviate the effects of interference in
multi-path routing. Our approach is based on two general principles. First, that
it is to work as a refinement phase over existing routing algorithms, thereby
inherently preserving, to the fullest possible extent, the advantages of any
given routing method. Second, that it is to rely only on information that is
locally available to the common origin of any given set of multiple paths
leading to the same destination. That is, only information that the origin can
obtain by communicating with its direct neighbors in the WMN should be used. One
intended consequence of the latter, in particular, is that refining the set of
paths departing from any common origin should be easily implementable by
straightforward message passing, and moreover, that any required calculation by
that node should be amenable to being carried out efficiently even if it
involves the solution of a computationally difficult problem. In order to comply
with these two principles, our approach operates on a previously established set
of paths leading from a common origin, say $i$, to a common destination, say
$j$. It operates exclusively on the neighborhood information stored at node $i$
itself or at any of its neighbors, say $k$, such that $k$ participates in some
of the $i$-to-$j$ paths, as well as on the information stored at these same
nodes regarding the routing of packets to node $j$. Once node $i$ has acquired
all this information, an undirected graph $G_{ij}$ is constructed that
represents every possible interference that can occur as packets get forwarded
toward $j$ by those of $i$'s neighbors that are on $i$-to-$j$ paths. Solving a
well-known NP-hard problem (that of finding a maximum weighted independent set)
on this typically small graph serves as a heuristic to decide which of the
$i$-to-$j$ paths to keep and which to discard.

It is important to note that, being determined with reference to graph $G_{ij}$,
the resulting set contains no two paths that interfere with each other as far as
node $i$'s neighbors are concerned, except of course for the inevitable
interference that may occur as packets leave $i$ or reach $j$. In our view, this
provides a sharp contrast between our approach and others that aim at weaker
forms of independence between the paths, for example by seeking paths that are
merely edge- or node-disjoint
\cite{Lee2001,Sung2001,Tsirigos2001,Cruz2003,Alicherry2006,Sheriff2006,Wang2006,Xiaojun2007,Wang2008,Wang2009}.
This is so because, as remarked elsewhere (e.g., \cite{Pearlman2000}),
independence by edge-disjointness encompasses independence by node-disjointness,
which in turn encompasses independence by noninterference. Of course, the
highest an independence relation's level in this hierarchy the easiest it is to
implement it as the multiple paths are discovered (not coincidentally, the
simple exchange of tokens between nodes suffices to produce edge- or
node-disjoint path sets \cite{Xuefei2004}).

We proceed in the following manner. First we state the problem in
graph-theoretic terms and give our solution in Section~\ref{sec:mra}. Then we
move, in Section~\ref{sec:methods}, to a presentation of the methodology we
followed in conducting our computational experiments. Our results are given in
Section~\ref{sec:results} and involve comparisons with some prominent routing
algorithms, viz.\ AODV \cite{Perkins1999}, AOMDV \cite{Marina2002}, OLSR
\cite{Jacquet2001}, and MP-OLSR \cite{Yi2011}. We used these algorithms both as
stand-alone methods and as bases to our own heuristic. Our results include
throughput and fairness \cite{Jain1998} comparisons based both on NS2.34
\cite{ns2} simulations and on the SERA link scheduling algorithm
\cite{Fabio2012}. We continue in Section~\ref{sec:discussion} with further
discussions and conclude in Section~\ref{sec:conclusion}.

\section{Problem formulation and heuristic}\label{sec:mra}

For $i$ and $j$ any two distinct nodes of the WMN, we begin by assuming that
some routing protocol already established a set $\mathcal{P}_{ij}$ of paths
directed from $i$ to $j$. Another key element is that, since we seek to
establish independence by noninterference, an interference model, along with its
assumptions, must be selected. Our choice is the protocol-based interference
model, together with the assumption that a node's communication and interference
radii are the same. Should a different model be selected or the two radii be
significantly different, the only effect would be for the graph construction
process outlined below, once adapted accordingly, to produce a different graph
(in particular, the interference radius could be chosen appropriately in order
for the protocol-based interference model to mimic the physical interference
model \cite{Shi2009}).

Under the assumptions of the protocol-based interference model, the
communication/interference radius is fixed at some value $R$, which we take to
be the same for all nodes. It follows that two nodes are neighbors of each other
in the WMN if and only if the Euclidean distance between them is no greater than
$R$. Moreover, since every link may transmit in both directions for error
control, it also follows that two links can interfere with each other if either
one's transmitter or receiver is a neighbor of the other's transmitter or
receiver \cite{Balakrishnan2004}. As will become apparent shortly, this has
important implications when modeling interference, since links that share no
nodes can still interfere with one another.

In keeping with the locality principle outlined in Section~\ref{sec:intro}, we
work on the premise that a node $k$'s knowledge is limited to the set $N_k$ of
its neighbors and the set $\mathrm{Next}_k(i,j)\subseteq N_k$ of neighbors to
which it may forward packets sent by node $i$ to node $j$ (assuming $k\neq j$).
Set $\mathrm{Next}_k(i,j)$, obviously, depends on the $j$-bound paths that leave
$i$ and go through node $k$. The problem we study is that of eliminating from
$\mathcal{P}_{ij}$ the fewest possible paths (in a weighted sense, to be
discussed later) so that the remaining path set, henceforth denoted by
$\mathcal{P}^\mathrm{R}_{ij}$, contains no mutually interfering paths except at
$i$ or $j$. However, owing once again to the issue of locality, we forgo both
optimality and feasibility a priori and settle for a heuristic instead. That is,
for the sake of locality we admit the possibility that, in the end, neither will
the selected paths be collectively optimal nor will the absence of interference
among them be guaranteed (except at those paths' second hops, which will be
non-interfering relative to one another necessarily).

Before proceeding, we tackle two special cases. The first one is that in which
$\mathcal{P}_{ij}$ contains the single-link path that connects $i$ directly to
$j$. In this case, we let $\mathcal{P}^\mathrm{R}_{ij}$ be the singleton that
contains that path, since no other arrangement can possibly do better. The
second special case is that in which $\vert\mathcal{P}_{ij}\vert=1$, provided
the single path contained in $\mathcal{P}_{ij}$ has at least two links, and is
motivated by the situations in which $\mathcal{P}_{ij}$ originates from
single-path routing. In this case, we enlarge $\mathcal{P}_{ij}$ before feeding
it to our path-selection heuristic. Letting node $k$ be such that
$\mathrm{Next}_i(i,j)=\{k\}$, we do this enlargement of $\mathcal{P}_{ij}$ by
including in it as many paths from $\mathcal{P}_{kj}$ as possible, each suitably
prefixed by the new origin $i$, provided none of the new paths is weightier than
the one initially in $\mathcal{P}_{ij}$. If enlargement turns out to be
impossible, then the problem becomes moot for $\mathcal{P}_{ij}$ and we let
$\mathcal{P}^\mathrm{R}_{ij}=\mathcal{P}_{ij}$.

We are then in position to introduce our refinement algorithm for multi-path
routing, henceforth referred to by the acronym MRA (for multi-path refinement
algorithm). The goal of MRA is to create an undirected graph $G_{ij}$
corresponding to the path set $\mathcal{P}_{ij}$ and to extract from it the
information necessary to determine $\mathcal{P}^\mathrm{R}_{ij}$. This graph's
node set, henceforth denoted by $V$, has one node for each of the paths in
$\mathcal{P}_{ij}$. Its edge set, denoted by $E$, is constructed in such a way
as to represent every interference possibility that can be inferred solely from
the sets $N_k$ and $\mathrm{Next}_k(i,j)$ for every $k\in N_i$ that participates
in at least one of the paths in $\mathcal{P}_{ij}$. Once graph $G_{ij}$ is
built, finding a maximum weighted independent set in it (i.e., a subset of $V$
containing no two nodes joined by an edge in $E$ and being as weighty as
possible) provides the best possible approximate decision on which of the paths
in $\mathcal{P}_{ij}$ should constitute $\mathcal{P}^\mathrm{R}_{ij}$, namely
those corresponding to the nodes in the maximum weighted independent set that
was found.

MRA proceeds as given next, following the introduction of some auxiliary
nomenclature. Given two neighboring WMN nodes, say $k$ and $k'$, we are
interested in the following two possibilities for the pair $k,k'$. A type-A pair
is part of at least one path in $\mathcal{P}_{ij}$ in such a way that
$k'\in\mathrm{Next}_k(i,j)$ with $k\in N_i$ or $k\in\mathrm{Next}_{k'}(i,j)$
with $k'\in N_i$. In a type-B pair, both $k$ and $k'$ belong to at least one
path in $\mathcal{P}_{ij}$ as well, but not the same path. Moreover, at least
one of them is a member of $N_i$ while the other, if not in $N_i$ as well, is in
$\mathrm{Next}_l(i,j)$ for some other $l\in N_i$. Note that type-A and -B pairs
constitute all the structural information that node $i$ can gather by strictly
local communication from its neighborhood $N_i$ in the WMN. In the case of a
type-A pair, we let $\mathrm{Paths}(k,k')$ be the set of $i$-to-$j$ paths to
which the pair belongs.

\begin{enumerate}
\item[(1)] Let $V$ have one node for each path in $\mathcal{P}_{ij}$. Add one
further node to $V$ for each type-B pair of neighboring WMN nodes. We refer to
these additional nodes as temporary nodes.
\item[(2)] Construct $E$ as follows:
\begin{enumerate}
\item[i.] Let $k,k'$ and $l,l'$ be two pairs of neighboring WMN nodes, each
pair being of type A or B. If it holds that $k=l$, $k=l'$, $k'=l$, or $k'=l'$,
then add an edge to $E$ between each node corresponding to a path in
$\mathrm{Paths}(k,k')$ (if the pair is of type A) or the corresponding temporary
node (if the pair is of type B) to each node corresponding to a path in
$\mathrm{Paths}(l,l')$ (if the pair is of type A) or the corresponding temporary
node (if the pair is of type B).
\item[ii.] Connect any two nodes in $V$ by an edge if, after the previous step,
the distance between them is $2$.
\item[iii.] Remove all temporary nodes from $V$ and all edges that touch them
from $E$.
\end{enumerate}
\item[(3)] Find a maximum weighted independent set of $G_{ij}$ and output
$\mathcal{P}^\mathrm{R}_{ij}$ accordingly.
\end{enumerate}

\begin{figure}[p]
\centering
\scalebox{0.75}{\includegraphics{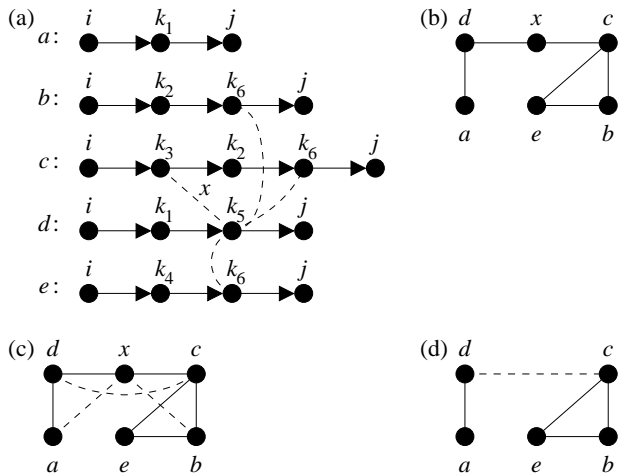}}
\caption{Construction of graph $G_{ij}$ for the path set
$\mathcal{P}_{ij}=\{a,b,c,d,e\}$. The paths in $\mathcal{P}_{ij}$ are shown in
panel (a), where each directed edge leads to a node in some $\mathrm{Next}(i,j)$
set (for example, $\mathrm{Next}_{k_1}(i,j)=\{j,k_5\}$) and each dashed edge
joins neighboring nodes that share none of the $i$-to-$j$ paths. We see in this
panel that $\mathrm{Paths}(k_2,k_6)=\{b,c\}$, and also that there are five
type-A node pairs ($k_1,j$; $k_2,k_6$; $k_2,k_3$; $k_1,k_5$; and $k_4,k_6$) as
well as one single type-B node pair ($k_3,k_5$, labeled $x$). Panel (b) shows
graph $G_{ij}$ as it stands after Steps~(1) and~(2).i, respectively for creating
its node set $V$ and initializing its edge set $E$ as a function of node
coincidence among all type-A or -B pairs. The distance-$2$ closure determined in
Step~(2).ii is shown in panel (c), with the additional edges represented as
dashed lines. Panel (d), finally, shows $G_{ij}$ as it stands at the end, after
Step~(2).iii. Notice, in particular, the fundamental role played by node $x$ in
identifying the interference between paths $c$ and $d$ under the protocol-based
interference model (it is through $x$, as the distance-$2$ closure is
determined, that $c$ and $d$ become connected). If equal weights were used for
all nodes, then clearly any of $\{a,b\}$, $\{a,c\}$, $\{a,e\}$, $\{b,d\}$, or
$\{d,e\}$ would qualify as a maximum weighted independent set of $G_{ij}$ in
Step~(3). However, using $1/C_p$ as the weight of the node corresponding to path
$p$ leads to $\{a,b\}$ and $\{a,e\}$ as the only possibilities, each of total
weight $1/2+1/3=5/6$, since $C_a=2$ and $C_b=C_e=3$. In the end, then, we have
either
$\mathcal{P}^\mathrm{R}_{ij}=\{a,b\}$ or $\mathcal{P}^\mathrm{R}_{ij}=\{a,e\}$,
each set containing paths whose second hops do not interfere with one another,
each set as large as possible in the weighted sense provided by the use of
$1/C_p$.}
\label{figure1}
\end{figure}

In these steps, graph $G_{ij}$ starts out as a graph with
$\vert\mathcal{P}_{ij}\vert$ nodes and gets enlarged by the addition of
temporary nodes that represent some of the possibilities of off-path
interference as nodes in $N_i$ engage in transmitting packets (Step~(1)). Then
it receives edges to account for the assumptions of the protocol-based
interference model (Steps~(2).i and~(2).ii) and is after that stripped of all
temporary nodes to end up with $\vert\mathcal{P}_{ij}\vert$ nodes once again
(Step~(2).iii). Step~(2).ii, in particular, accounts for interference in the
WMN when a link's transmitter or receiver does not coincide with (but is a
neighbor of) another link's transmitter or receiver. The last MRA step,
Step~(3), is the determination of a maximum weighted independent set of
$G_{ij}$. We use node weights such that, for the node corresponding to path
$p\in\mathcal{P}_{ij}$, the weight is $1/C_p$, where $C_p$ is the path's hop
count. In other words, shorter paths tend to be favored over longer ones as
$\mathcal{P}^\mathrm{R}_{ij}$ is extracted from $\mathcal{P}_{ij}$. Clearly,
though, any other desired criterion can be used as well. An illustration of how
MRA works is given in Fig.~\ref{figure1}.

\section{Methods}\label{sec:methods}

We evaluated the performance of MRA through extensive experimentation with the
following routing algorithms: AODV \cite{Perkins1999}, AOMDV \cite{Marina2002},
OLSR \cite{Jacquet2001}, and MP-OLSR \cite{Yi2011}. For the purpose of
conciseness, we henceforth refer to the combination of each of these algorithms
with MRA as R-AODV, R-AOMDV, R-OLSR, and R-MP-OLSR, respectively. We remark
that, since both AODV and OLSR are single-path routing algorithms, handling the
paths they generate falls into one of the special cases discussed in
Section~\ref{sec:mra}. Our experiments were run in the network simulator NS2.34
(NS2 henceforth) \cite{ns2} and in a simulator that employs the SERA link
scheduling algorithm \cite{Fabio2012}, briefly described later in this section.
We used two different configurations of routing-algorithm parameters, and
likewise two different configurations of NS2 parameters (one for the
path-discovery process and another for performance evaluation). These
configurations were selected during initial tuning experiments and will be
presented shortly.

\subsection{Topology generation}

We generated four types of network according to the maximum number of neighbors,
$\Delta$, a node may have.  Each network was generated by placing $n$ nodes
inside a square of side $1500$. The first of these nodes was positioned at the
center of the square and the remaining nodes were placed randomly as a function
of the communication/interference radius $R$ introduced at the beginning of
Section~\ref{sec:mra}. Their placement was subject to the constraints that each
node would have at least one neighbor, that no node would be closer to any other
than $25$ units of Euclidean distance, and that no node would have more than
$\Delta$ neighbors. No more than $1000$ attempts at positioning nodes were
allowed; if this limit was reached then the growing network was discarded and
the generation of a new one was started. The value of $R$ was determined so
that the expected density of nodes inside a radius-$R$ circle would be
proportional to $\Delta/R^2$ (assuming some uniformly random form of placement),
and be moreover about the same density as that of the whole network. It follows
that $\Delta/R^2 \propto n$. We chose the proportionality constant to yield
$R=200$ for $n=80$ and $\Delta=4$, whence $R=200\sqrt{20\Delta/n}$. Of all the
networks generated, there are $100$ networks for each combination of
$n\in\{60,80,100,120\}$ and $\Delta\in\{4,8,16,32\}$, thus totaling $1600$
networks.

\subsection{Path discovery}

For each of the $1600$ networks we randomly generated $100$ sets of node pairs
to function as origin-destination pairs (instances of the $i,j$ pair we have
been using throughout). Each set comprises $n$ pairs and no node was allowed to
appear more than once in any set as an origin. For each of the node-pair sets
and each of the four routing algorithms (AODV, AOMDV, OLSR, and MP-OLSR) we
obtained $n$ path sets (instances of $\mathcal{P}_{ij}$). Likewise, for each of
the node-pair sets and each of the four refined routing algorithms (R-AODV,
R-AOMDV, R-OLSR, and R-MP-OLSR) we obtained another $n$ path sets (instances of
$\mathcal{P}^\mathrm{R}_{ij}$).

For each routing algorithm, the discovery of each $\mathcal{P}_{ij}$ instance
(i.e., for a single origin and a single destination) proceeded as follows. After
loading the network topology onto NS2 we conducted a $15$-second simulation with
one flow agent for the single origin $i$ and the single destination $j$, using a
CBR of one $1000$-byte packet per second. The remaining pairs were handled
likewise after resetting the simulator. We remark that this one-pair-at-a-time
strategy, as opposed to generating paths for all $n$ pairs in the same set
concomitantly, was meant to minimize packet loss due to path overload and also
to avoid the possible interference of a previously discovered path with the
discovery of a new one. Of course, this argument is only valid for on-demand
routing algorithms (AODV and its variants depend on network load, while the OLSR
variants always find identical paths for the same network topology), but we
proceeded in this way in all cases. As a consequence, our experiments are
entirely reproducible.

We set NS2 to its default configuration, but employed the DRAND MAC protocol
\cite{Rhee2006} to avoid collisions in the path-discovery process. To adjust the
radius $R$ we also set the parameter \emph{RXThresh\_} (RXT) to the appropriate
value given by the program \emph{threshold.cc} (cf.\ the NS2 manual). We used
the implementations of the AODV, OLSR, and AOMDV routing agents available in
version 2.34 of NS2 and the MP-OLSR routing agent available at \cite{mpolsr}.
For AOMDV, we made the small modifications proposed by \cite{YuHua2005} to
discover only node-disjoint paths with at most $K$ paths for each
origin-destination pair of nodes. We adopted these modifications because
node-disjoint paths are clearly more interference-free than otherwise. We chose
$K=5$ because it achieved the best throughput values for $2\le K\le 7$. The same
modifications were effected on MP-OLSR (as proposed by \cite{Xun2005}). Out of
the same range for $K$, and for the same reason as above, we used $K=3$ for
$\Delta\in\{4,8\}$ and $K=5$ for $\Delta \in \{16,32\}$.

\subsection{Performance evaluation}

Once we fix a value for $n$ and a value for $\Delta$, there are $10^4$ path sets
on which to evaluate the performance of MRA. Each of these sets is relative to
$n$ origin-destination pairs. In our experiments, we randomly grouped these
pairs into $n$ sets, each containing a different number of origin-destination
pairs (i.e., one set containing a single pair, another containing two pairs, and
so on), and simulated the network's behavior in transporting predominantly heavy
traffic from the origins to the destinations. We use $OD$ to denote the set of
pairs in question, hence $1\le\vert OD\vert\le n$, and let
$\mathcal{P}=\bigcup_{ij\in OD}\mathcal{P}_{ij}$ and
$\mathcal{P}^\mathrm{R}=\bigcup_{ij\in OD}\mathcal{P}^\mathrm{R}_{ij}$. For the
sake of normalization, all our performance results are presented against the
pair density $\theta=\vert OD\vert/n\in(0,1]$. 

Each experiment began by loading the corresponding network onto NS2, with MAC
set to 802.11 and the routing agent to NOAH \cite{noah}. NOAH works only with
fixed paths that have to be configured manually and therefore does not send
routing-related packets, thus providing the ideal setting for a performance
evaluation free of any interference from control packets. Next we started a CBR
traffic flow from each origin to each destination and measured the number of
successfully delivered packets during the last $120$ of the $135$ seconds of
simulation (following, therefore, a warm-up period of $15$ seconds).

During the initial, tuning experiments we varied three of the NS2 parameters
widely. These were the carrier sense threshold (CST), aiming to increase the
spatial reuse and consequently the throughput \cite{Kim2006}, and the CBR
parameter as well as the network transmission rate, aiming to obtain as much
throughput and fairness as possible (cf.\ Section \ref{subsec:theresults}).
Table~\ref{table1} presents the parameter ranges of our tuning experiments,
during which the highest CBR rate that afforded some gain in throughput was
identified for each value of $\Delta$, and also lowest rate that afforded some
gain in fairness.\footnote{We observed in NS2 simulations that, within limits,
increasing the CBR \emph{rate\_} parameter tends to lead to an increase in
throughput while decreasing it tends to lead to an increase in fairness.} Our
choices for use thereafter while conducting performance evaluation are shown in
Table~\ref{table2}, where CBR1 and CBR2 refer to such highest and lowest rate,
respectively.

\begin{table}[t]
\centering
\caption{Parameters used for tuning.}
\begin{tabular}{ccc}
\hline
NS2 parameter  & Interval                                    & Increment \\ \hline
CBR interval   & $0.001\textrm{s}\twoldots 0.005\textrm{s}$  & 0.0001    \\
CST            & $0.1\textrm{RXT}\twoldots 2\textrm{RXT}$    & 0.1RXT    \\
Data/basic rate& 1/1, 2/1, 11/2 Mbps                         &           \\
\end{tabular}

\label{table1}
\end{table}

\begin{table}[t]
\centering
\caption{Parameters used for performance evaluation.}
\begin{tabular}{ccccc}
\hline
NS2 parameter   & \multicolumn{4}{c}{$\Delta$}           \\ \cline{2-5}
                & 4       & 8       & 16      & 32       \\ \hline 
CBR1 interval   & 0.0025s & 0.0027s & 0.0029s & 0.0031s  \\  
CBR2 interval   & \multicolumn{4}{c}{0.0045s}             \\
CBR packet size & \multicolumn{4}{c}{1000 bytes}         \\
CST             & 0.6RXT  & 0.7RXT  & 0.8RXT  & 0.9RXT   \\
Data/basic rate & \multicolumn{4}{c}{11/2 Mbps}          \\
\end{tabular}

\label{table2}
\end{table}

All our NS2 experiments were carried out under 802.11, which is a CSMA protocol.
As an alternative setting that might provide some insight into the performance
of MRA under some TDMA scheme (an approach fundamentally distinct from CSMA
\cite{Gummalla2000}), we selected the SERA link scheduling algorithm
\cite{Fabio2012}. SERA seeks to schedule the links of a set of paths while
striving to maximize throughput on those paths. It is therefore quite well
suited to the task at hand. The throughput that our SERA simulator provides is
given in terms of time slots, so in order to achieve a meaningful basis for
comparing CSMA- and TDMA-based results a translation is needed of such
throughput figures into those provided by NS2 in the CSMA experiments. We did
this by resorting to a very simple NS2 simulation to determine the duration of
a time slot. In this experiment, a node sends packets to another and the time
for successful deliveries is recorded. Since this time is conceptually the same
that in SERA is taken to be a time slot, the translation from one setting to the
other can be accomplished easily. Using the parameter values shown in
Table~\ref{table2}, we found a time-slot duration of $0.002$ seconds. This is
the duration we use, together with the ND-BF numbering scheme for SERA and its
$B$ parameter set to $2$ (cf.\ \cite{Fabio2012}).

\section{Computational results}\label{sec:results}

We divide our results into two categories. First we present a statistical
analysis of the networks generated and the path sets obtained by the original
routing algorithms and by their refinements through the use of MRA. Then we
present the ratios of the refined algorithms' throughputs to those of their
corresponding originals (absolute values are given in
Figs.~\ref{figureS1}--\ref{figureS3}) and also fairness figures. For the purpose
of conciseness we report only on the $n=120$ results, since they are
qualitatively similar to those related to the other three values of $n$ we used.
In Section~\ref{sec:discussion}, though, we do discuss some of the quantitative
differences that were observed.

\subsection{Properties of the networks generated}

The $1600$ networks we generated are that same that were used in
\cite{Fabio2012}. We refer the reader to Section~8.1 of that publication for a
variety of the networks' statistical properties, such as the occurrence of
topologies structured in some particular way and some of their
structure-related distributions. Here we concentrate on presenting those
properties that pertain to routing both before and after refinement, since they
are the ones we have found useful in helping explain the throughput results we
present later.

The average path multiplicity (number of paths) per origin-destination pair in
the path sets $\mathcal{P}$ and $\mathcal{P}^\mathrm{R}$ is given in
Table~\ref{table3} for every combination of $\Delta$ and $\vert OD\vert$.
Note that MP-OLSR is absent from the table in spite of being a multi-path
algorithm, the reason being that in this case the $K$ parameter does not work
as an upper bound (as it does for AOMDV), but rather as the fixed number of
paths to be found. We observe in the table that the average path multiplicity
increases monotonically with $\Delta$ for fixed $\vert OD\vert$, which is
expected from the well-known fact that the number of possible paths between the
same two nodes grows with $\Delta$ in arbitrary graphs \cite{Hoffman1963}. On
the other hand, increasing $\vert OD\vert$ for fixed $\Delta$ causes very little
variation, probably owing to the method we used to discover the paths in the
first place (i.e., by handling each origin-destination pair independently of the
others). The overall pruning effect of MRA is also clearly manifest in the
table, since in all cases the average for an algorithm's refined version is less
than that of the original algorithm (i.e., on average we have
$\vert\mathcal{P}^\mathrm{R}\vert<\vert\mathcal{P}\vert$ for all routing
algorithms).

\begin{table}[t]
\centering
\caption{Average path multiplicity per origin-destination pair for $n=120$. Data
are averages over $10^4$ instances of $\mathcal{P}$ or $\mathcal{P}^\mathrm{R}$
for each combination of $\Delta$ and $\vert OD\vert$.}
\small
\begin{tabular}{ccccccc}
\hline
$\Delta$ & $\vert OD\vert $ & R-AODV & AOMDV & R-AOMDV & R-OLSR & R-MP-OLSR \\ \hline
4        & $1\twoldots 10$   & 1.5    & 3.3   & 1.7     & 1.4    & 1.5 \\
         & $11\twoldots 90$  & 1.5    & 3.4   & 1.8     & 1.5    & 1.6 \\
         & $91\twoldots 120$ & 1.6    & 3.4   & 1.8     & 1.5    & 1.6 \\ \hline
8        & $1\twoldots 10$   & 1.6    & 3.6   & 1.9     & 1.6    & 1.7 \\
         & $11\twoldots 20$  & 1.7    & 3.6   & 2.1     & 1.6    & 1.7 \\
         & $21\twoldots 70$  & 1.7    & 3.7   & 2.1     & 1.6    & 1.7 \\
         & $71\twoldots 80$  & 1.7    & 3.7   & 2.1     & 1.6    & 1.8 \\
         & $81\twoldots 120$ & 1.7    & 3.7   & 2.2     & 1.6    & 1.8 \\ \hline
16       & $1\twoldots 10$   & 2      & 3.9   & 2.9     & 2      & 2   \\
         & $11\twoldots 50$  & 2.1    & 3.9   & 2.9     & 2      & 2   \\
         & $51\twoldots 90$  & 2.1    & 4     & 3       & 2.1    & 2   \\
         & $91\twoldots 120$ & 2.1    & 4     & 3       & 2.1    & 2.1 \\ \hline
32       & $1\twoldots 10$   & 2.2    & 4.1   & 3.1     & 2.2    & 2.2 \\
         & $11\twoldots 60$  & 2.2    & 4.1   & 3.1     & 2.2    & 2.3 \\
         & $61\twoldots 70$  & 2.3    & 4.2   & 3.2     & 2.2    & 2.4 \\
         & $71\twoldots 80$  & 2.3    & 4.2   & 3.2     & 2.2    & 2.5 \\
         & $81\twoldots 90$  & 2.3    & 4.2   & 3.2     & 2.2    & 2.6 \\
         & $91\twoldots 120$ & 2.3    & 4.2   & 3.2     & 2.3    & 2.7 \\ \hline
\end{tabular}

\label{table3}
\end{table}

\begin{figure}[p]
\centering
\scalebox{0.55}{\includegraphics{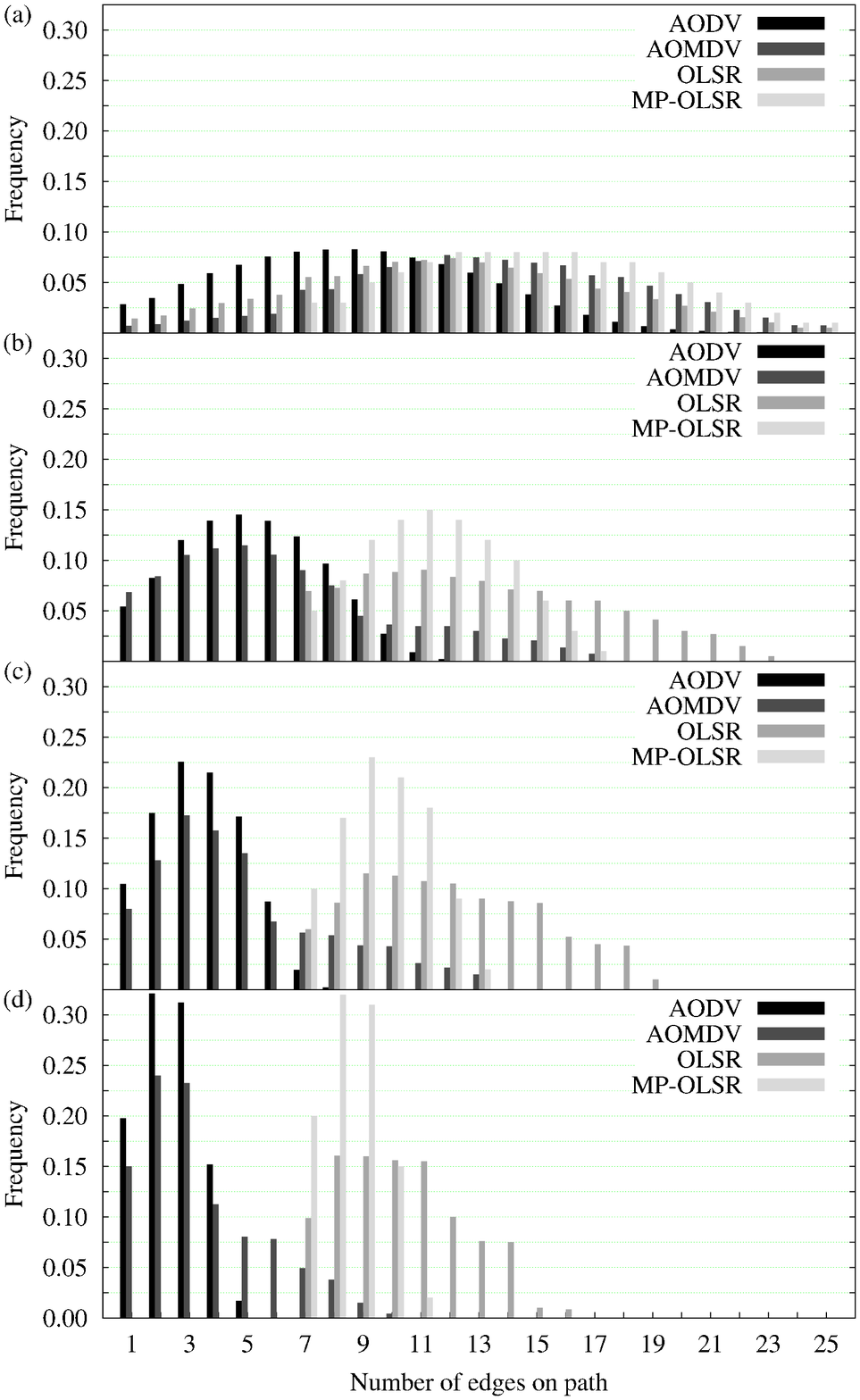}}
\caption{Distribution of path sizes for the original algorithms with $n=120$,
for $\Delta=4$ (a), $\Delta=8$ (b), $\Delta=16$ (c), and $\Delta=32$ (d). For
each value of $\Delta$ the distribution refers to $100$ networks and $100$ path
sets per network, each corresponding to $n$ origin-destination pairs.}
\label{figure2}
\end{figure}

\begin{figure}[p]
\centering
\scalebox{0.55}{\includegraphics{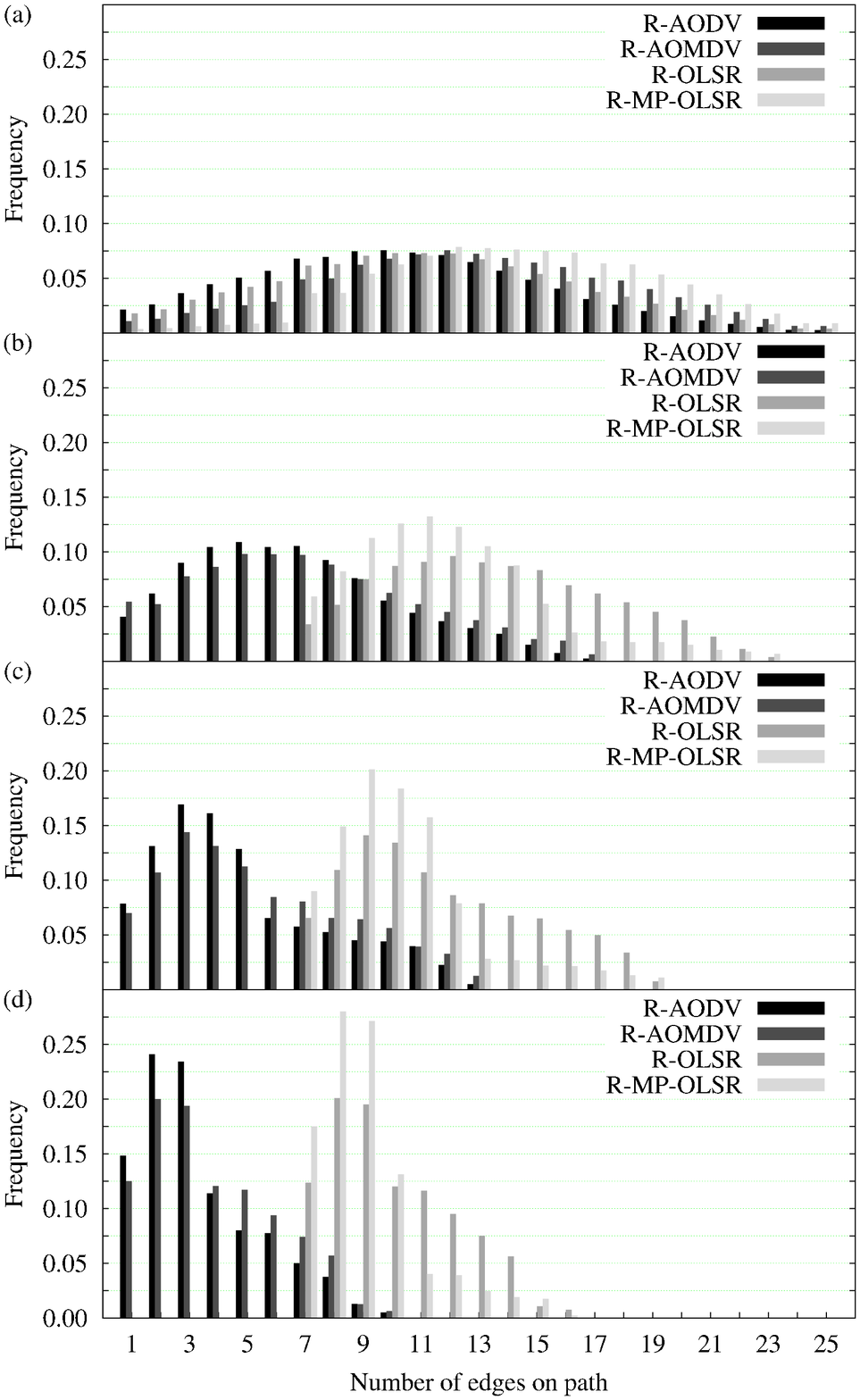}}
\caption{Distribution of path sizes for the refined algorithms with $n=120$, for
$\Delta=4$ (a), $\Delta=8$ (b), $\Delta=16$ (c), and $\Delta=32$ (d). For each
value of $\Delta$ the distribution refers to $100$ networks and $100$ path sets
per network, each corresponding to $n$ origin-destination pairs.}
\label{figure3}
\end{figure}

The path-size distributions for $\mathcal{P}$ and $\mathcal{P}^\mathrm{R}$ with
$\vert OD\vert=n$ (that is, for the original algorithms and their refinements,
using the full path sets, as generated) are given in Figs.~\ref{figure2}
and~\ref{figure3}, respectively, for every value of $\Delta$. Note first that,
as expected by virtue of the well-known dependency of path sizes on the average
degree of nodes \cite{Dirac1952}, increasing $\Delta$ leads the path-size
distribution for a given algorithm to peak at ever smaller values. Another
expected result is that the OLSR variants all allow for longer paths than those
of AODV. The reason for this is that the multi-point relay (MPR) concept at the
heart of OLSR tends to cause longer paths to be produced than the shortest-path
algorithm (SPA) used by AODV. It is also noteworthy that the application of MRA
to yield the refined algorithms has no impact on the distributions other than
causing some of the greatest path sizes to occur more frequently.

\subsection{Results}\label{subsec:theresults}

\begin{figure}[p]
\centering
\scalebox{0.55}{\includegraphics{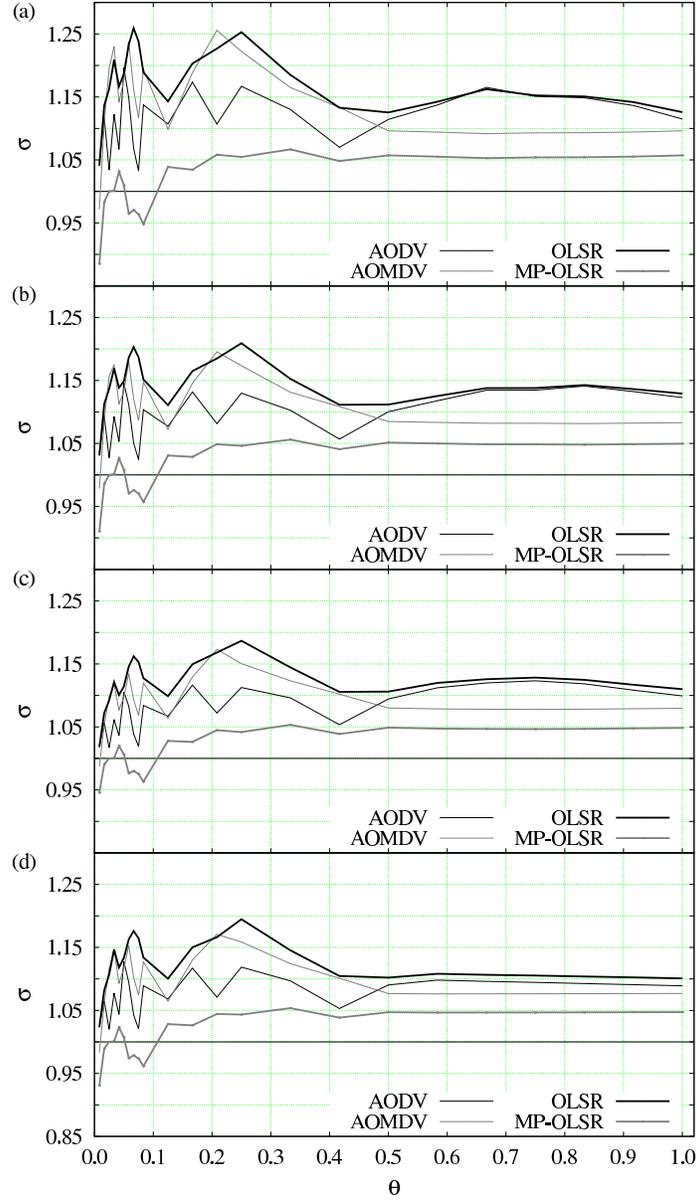}}
\caption{Throughput ratio for CBR1 with $n=120$, for $\Delta=4$ (a), $\Delta=8$
(b), $\Delta=16$ (c), and $\Delta=32$ (d). Data are averages over the $10^4$
path sets that correspond to each value of $\Delta$ for each value of $\theta$.
Confidence intervals are less than $1\%$ of the mean at the $95\%$ level, so
error bars are omitted.}
\label{figure4}
\end{figure}

\begin{figure}[p]
\centering
\scalebox{0.55}{\includegraphics{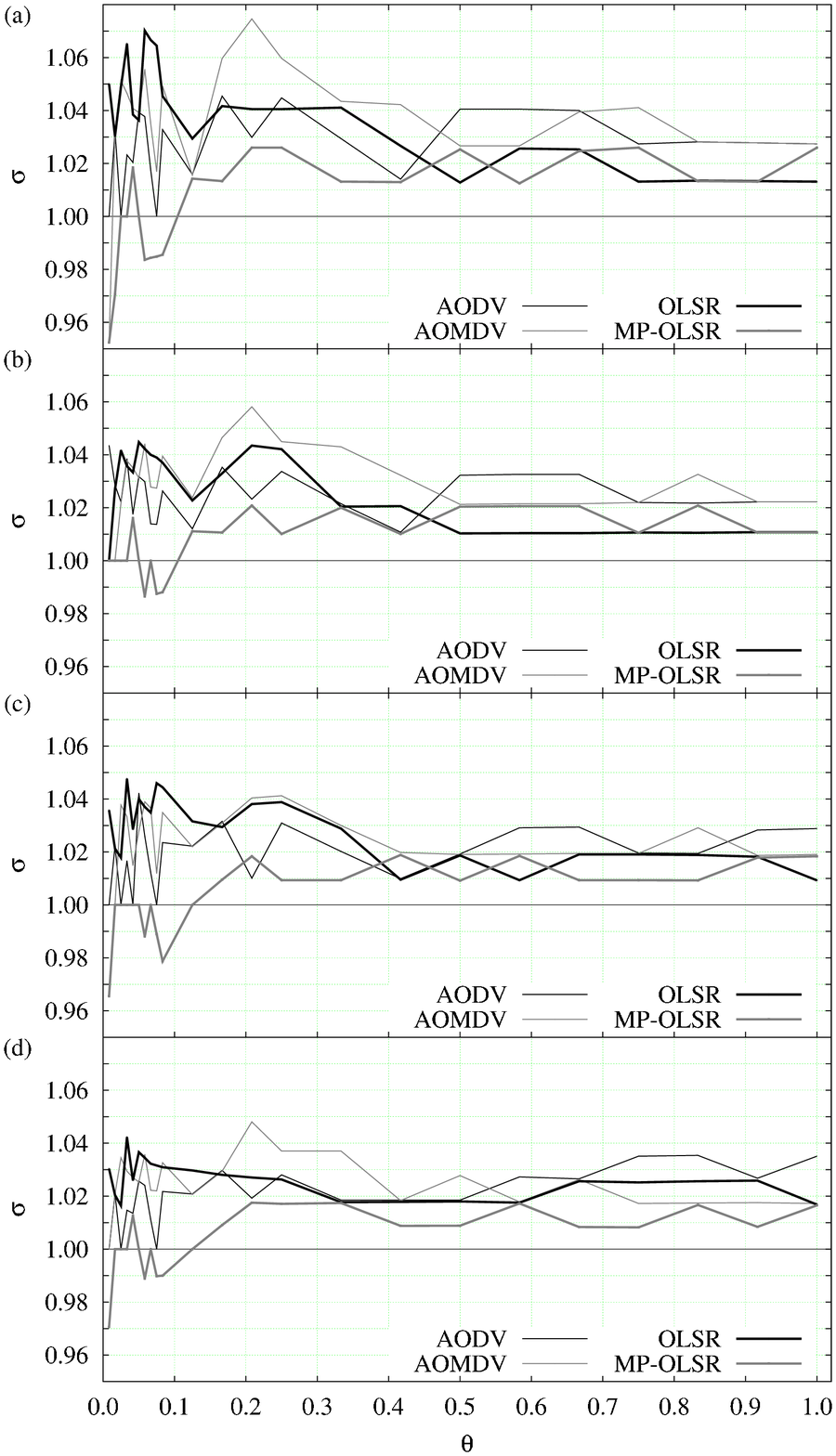}}
\caption{Throughput ratio for CBR2 with $n=120$, for $\Delta=4$ (a), $\Delta=8$
(b), $\Delta=16$ (c), and $\Delta=32$ (d). Data are averages over the $10^4$
path sets that correspond to each value of $\Delta$ for each value of $\theta$.
Confidence intervals are less than $1\%$ of the mean at the $95\%$ level, so
error bars are omitted.}
\label{figure5}
\end{figure}

\begin{figure}[p]
\centering
\scalebox{0.55}{\includegraphics{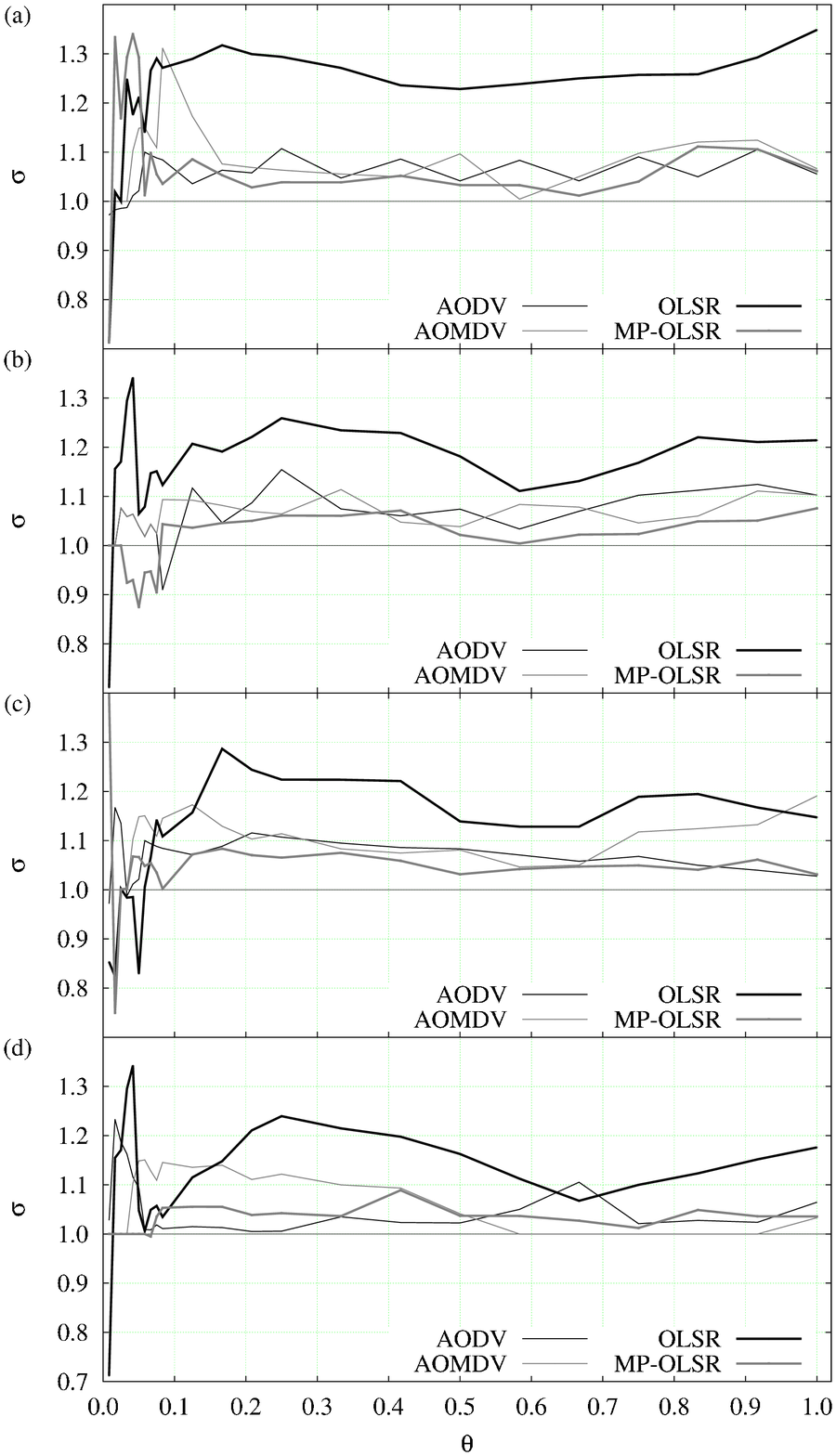}}
\caption{Throughput ratio for SERA with $n=120$, for $\Delta=4$ (a), $\Delta=8$
(b), $\Delta=16$ (c), and $\Delta=32$ (d). Data are averages over the $10^4$
path sets that correspond to each value of $\Delta$ for each value of $\theta$.
Confidence intervals are less than $1\%$ of the mean at the $95\%$ level, so
error bars are omitted.}
\label{figure6}
\end{figure}

Our computational results relating to throughput are summarized in
Fig.~\ref{figure4} for CBR1, Fig.~\ref{figure5} for CBR2, and Fig.~\ref{figure6}
for SERA. Each figure is organized as a set of four panels, each for one of the
four values of $\Delta$. All three figures show the behavior of the ratio, here
denoted by $\sigma$, of each refined algorithm's throughput to that of its
original version. All plots are given against the ratio $\theta$ introduced
earlier, which indicates what fraction of the $n$ origin-destination pairs
corresponding to each path set is being taken into account. In all panels a
highlighted horizontal line is used to mark the $\sigma=1$ threshold, above
which a refined algorithm can be said to have overtaken its original version.

Except for a few cases of $\theta<0.1$ (in which only a few origin-destination
pairs coexist in the network and therefore refinement to achieve
path-independence by noninterference is probably pointless to begin with), it
follows from Figs.~\ref{figure4}--\ref{figure6} that MRA was effective to some
extent in all cases. Fixing the simulation scenario (CBR1, CBR2, or SERA)
reveals that the behavior of $\sigma$ depends very little on the value of
$\Delta$ or $\theta$ (provided $\theta$ is sufficiently large). Overall, the
values of $\sigma$ seem best for the coupling of OLSR with SERA, followed by
CBR1 coupled with either OLSR or MP-OLSR, and lastly for CBR2 without any marked
preference for any routing method. Recall that the SERA scenario is TDMA-based,
while both CBR1 and CBR2 are CSMA-based, with CBR1 operating at the higher
rates.

Another perspective from which it is worth examining performance is that of
fairness in the distribution of traffic through the paths. In other words, given
an origin-destination pair and the multiple paths leading from the origin to the
destination, we look at how traffic gets distributed through the various paths.
One way of quantifying this is by means of the fairness index \cite{Jain1998}.
Given a set of paths $\mathcal{Q}$,\footnote{We use $\mathcal{Q}$ as a place
holder for either $\mathcal{P}$ or $\mathcal{P}^\mathrm{R}$, depending on
whether the routing algorithm in question is one of the originals or one of the
refinements through MRA.} the corresponding fairness index can be defined as
$(\sum_{p\in\mathcal{Q}}x_p)^2/\vert\mathcal{Q}\vert\sum_{p\in\mathcal{Q}}x_p^2$,
where $x_p$ is the number of packets delivered to $j$ through path $p$ during
the experiment. The fairness index ranges from $1/\vert\mathcal{Q}\vert$ to $1$,
indicating when equal to $1$ that traffic is evenly distributed among the paths.

\begin{figure}[p]
\centering
\scalebox{0.55}{\includegraphics{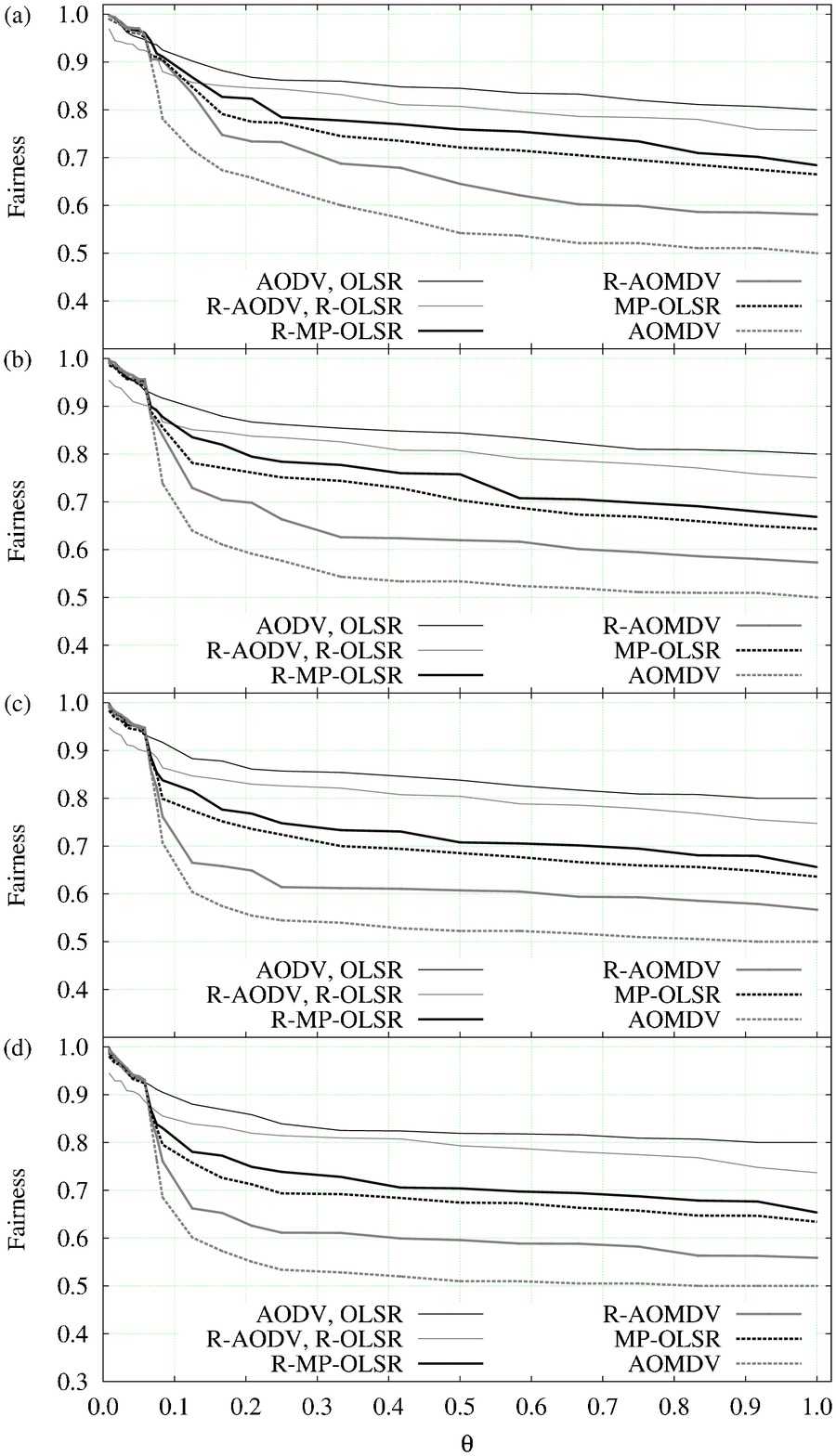}}
\caption{Fairness index for CBR1 with $n=120$, for $\Delta=4$ (a), $\Delta=8$
(b), $\Delta=16$ (c), and $\Delta=32$ (d). Data are averages over the $10^4$
path sets that correspond to each value of $\Delta$ for each value of $\theta$.
Confidence intervals are less than $1\%$ of the mean at the $95\%$ level, so
error bars are omitted.}
\label{figure7}
\end{figure}

\begin{figure}[p]
\centering
\scalebox{0.55}{\includegraphics{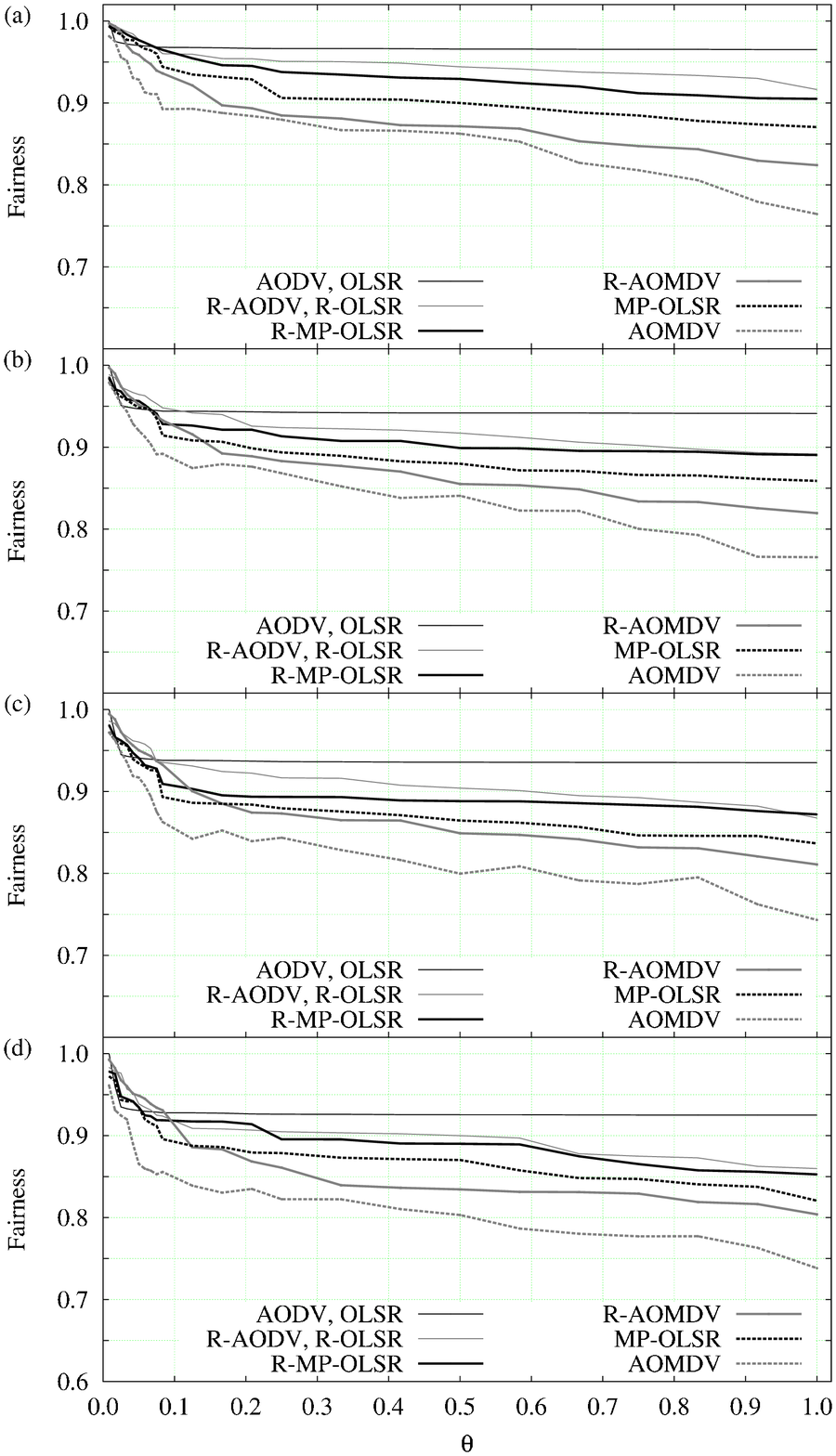}}
\caption{Fairness index for CBR2 with $n=120$, for $\Delta=4$ (a), $\Delta=8$
(b), $\Delta=16$ (c), and $\Delta=32$ (d). Data are averages over the $10^4$
path sets that correspond to each value of $\Delta$ for each value of $\theta$.
Confidence intervals are less than $1\%$ of the mean at the $95\%$ level, so
error bars are omitted.}
\label{figure8}
\end{figure}

\begin{figure}[p]
\centering
\scalebox{0.55}{\includegraphics{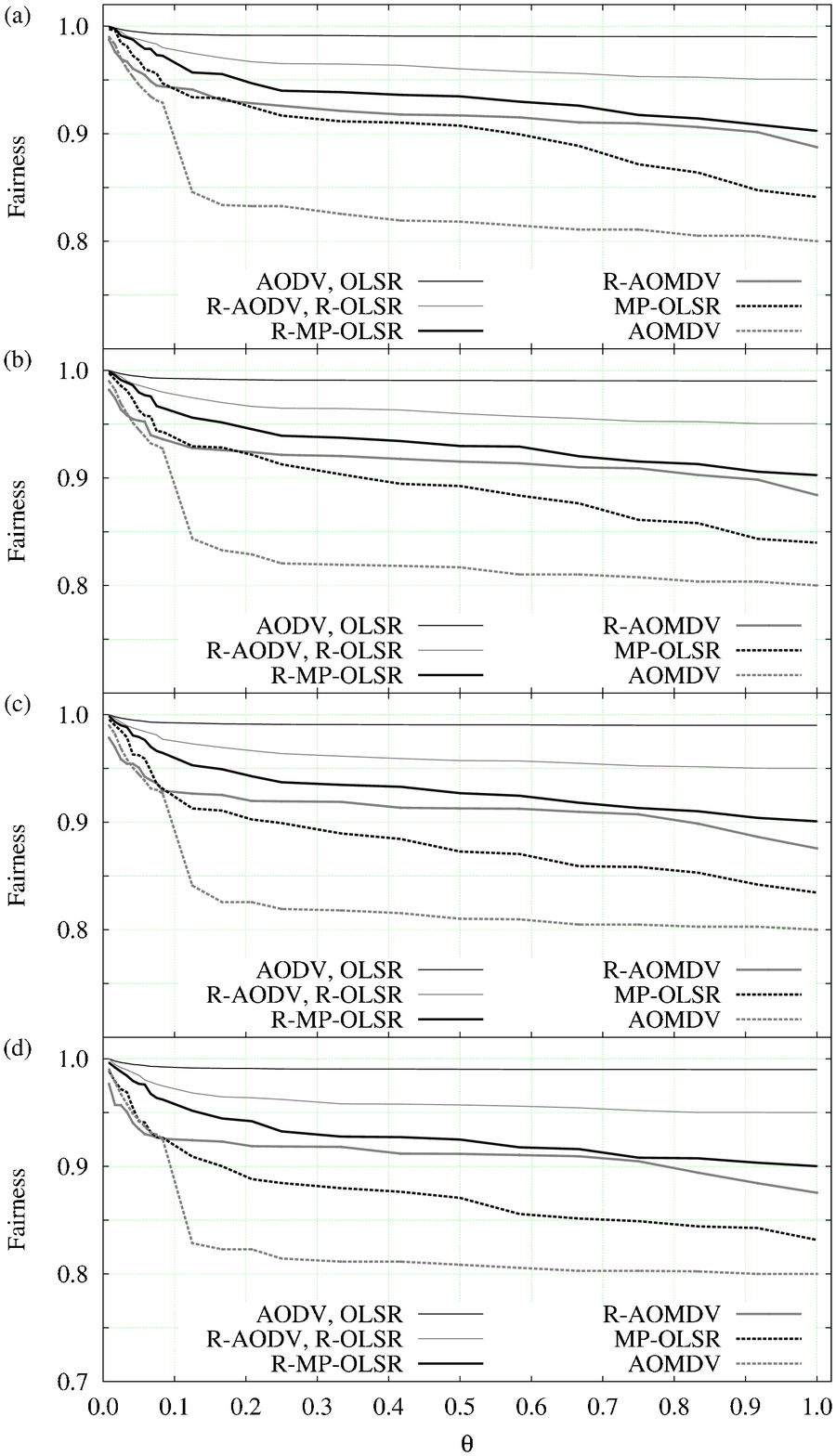}}
\caption{Fairness index for SERA with $n=120$, for $\Delta=4$ (a), $\Delta=8$
(b), $\Delta=16$ (c), and $\Delta=32$ (d). Data are averages over the $10^4$
path sets that correspond to each value of $\Delta$ for each value of $\theta$.
Confidence intervals are less than $1\%$ of the mean at the $95\%$ level, so
error bars are omitted.}
\label{figure9}
\end{figure}

We give results on the fairness index in Figs.~\ref{figure7}--\ref{figure9},
respectively for CBR1, CBR2, and SERA. Note initially that, somewhat
unexpectedly (owing to the algorithms' markedly different strategies), AODV and
OLSR are statistically indistinguishable from each other as far as fairness is
concerned. The same holds for their refined versions, respectively R-AODV and
R-OLSR. Note also that the fairness index, in all cases, tends to decrease as
$\theta$ is increased. This means that, as might be expected, the presence of
denser end-to-end traffic tends to disrupt the balance between paths more
easily. AODV and OLSR have the best figures overall, better even than their
refined versions. So, unlike throughput, fairness does not seem to improve as we
extend a single-path algorithm's set of paths and then apply MRA for refinement.
All multi-path strategies, on the other hand, can be seen to benefit from the
use of MRA.

Another interesting trend that can be observed in
Figs.~\ref{figure7}--\ref{figure9} is that, though always decaying with $\theta$
as we noted, in general the fairness index is best for SERA, followed by CBR2,
then by CBR1. While we believe the position of SERA in this rank to be closely
related to its TDMA-based nature, it is curious to observe that the relative
positions of CBR1 and CBR2 are exchanged with respect to what we observed for
throughput. It seems, then, that in selecting between the rates associated with
CBR1 and CBR2 one is automatically forced to favor throughput over fairness or
conversely.

Our definition of the fairness index given above is only one of the
possibilities in the context of multi-path routing. Another alternative is to
coalesce all packets delivered from one origin, say $i$, to one destination, say
$j$, into a single number $x_{ij}$ and then compute the fairness index as
$(\sum_{ij\in OD}x_{ij})^2/\vert OD\vert\sum_{ij\in OD}x_{ij}^2$. Proceeding in
this way would shift the focus of the fairness index from paths to
origin-destination pairs. We give no detailed results on this alternative, but
do provide an example with SERA in Fig.~\ref{figureS4}. There is clearly great
similarity with the results in Fig.~\ref{figure9}, but the numbers tend to be
higher.

\section{Discussion}\label{sec:discussion}

As we remarked earlier, we have provided no results for $n\in\{60,80,100\}$
because, essentially, they are indistinguishable from those of $n=120$ in
qualitative terms. We do remark, however, that a few noteworthy quantitative
differences were observed. For example, in the case of SERA a higher throughput
ratio $\sigma$ was sometimes observed for reduced $n$ but the same value of the
pair density $\theta$. This stems not only from the fact that the average path
size decreases as $n$ decreases, but more generally from the fact that the
number of links for the same $\theta$ is lower in the smaller networks, thus
fewer links interfere with one another and fewer links have to be scheduled.
Such improvement in the value of $\sigma$, therefore, seems to depend on the
network's path-size distribution.

As we noted briefly in Section~\ref{subsec:theresults}, often OLSR and MP-OLSR
turned out to be the routing methods most prone to benefit from the refinement
provided by MRA. One clue as to why this is so may be already present in
Table~\ref{table3}, where the refined versions of these two methods have some
of the lowest path multiplicities overall, thence a tendency to incur less
interference. However, this holds also for the refined version of AODV, so there
has to be some other distinguishing aspect. While our data are not sufficient to
provide a definitive answer, we believe the two methods' superiority to be owed
to a combination of MRA (which attempts to reduce path multiplicity to lower
interference) with the MPR concept that is intrinsic to the OLSR variants (which
in turn attempts to provide an initial set of paths as spatially distributed as
possible). As for AODV, the SPA at its core probably produces paths that are
less spatially separated \cite{Jiazi2008,Biradar2010}.

In a related vein, the results of Section~\ref{subsec:theresults} also point at
SERA as providing superior throughput results vis-\`{a}-vis those of CBR1 and
CBR2, and similarly for CBR1 with respect to CBR2. Because such results refer to
throughput gains after refinement by MRA, more detailed data are needed for a
direct comparison of throughputs. These are shown in
Figs.~\ref{figureS5}--\ref{figureS7}, which clearly confirm the rank. Once again
the reason for this is not totally clear, but it may be a manifestation of
SERA's properties rather than of the superiority of its underlying TDMA scheme
over the CSMA protocols of CBR1 and CBR2. After all, a considerable amount of
research has been directed toward the TDMA versus CSMA question, without however
reaching an agreement
\cite{Ding2002,Gupta2007,Ashutosh2009,Banaouas2009}.

Another curiosity related to this issue of how SERA compares to CBR1 and CBR2
has to do with the total absence of traffic on some paths. While SERA precludes
this from occurring as a matter of design principle, unused paths do occur in
the other two cases. During the corresponding NS2 simulations this persisted
even if simulation times were extended or the synchronization of multiple CBRs
was reordered or entirely removed.\footnote{As these appear to be commonly
occurring problems of NS2 agents.} As it turns out, it seems that certain paths
remained unused so that throughput could be increased on other paths. Data on
the most critical cases, viz.\ those in which no packets at all were delivered
for an origin-destination pair, are shown in Figs.~\ref{figureS8}
and~\ref{figureS9}. Examining these data confirms our expectation that the
single-path algorithms should be less prone to the occurrence of such extreme
cases. It also reveals that both R-AOMDV and R-MP-OLSR had fewer such
occurrences than their corresponding originals (i.e., MRA seems to have
attenuated the problem).

\section{Concluding remarks}\label{sec:conclusion}

MRA is a heuristic for the refinement of routing paths in WMNs. It was developed
with multi-path routing algorithms in mind but works also on single-path
algorithms (through a manipulation of its input to obtain multiple paths from
the overall set of single paths). MRA is fully local, in the sense that it
depends only on information that is readily available to each node and its
immediate neighborhood in the network. Being local means that the extra control
traffic it may entail is negligible, especially if we consider that it need
happen only once for a fixed set of paths. It involves the solution of an
NP-hard problem, that of finding a maximum weighted independent set in a graph,
but the inputs involved are typically very small, leading to negligible running
times if compared to arbitrary graphs \cite{Pardalos1991}.

Our computational experiments with MRA on a TDMA setting (SERA) and two CSMA
settings (CBR1 and CBR2) revealed improvements in throughput of up to $30\%$ for
both the AODV and OLSR routing methods and their multi-path variants. The latter
were also improved by MRA in terms of fairness. Finding out just how close these
improvements come to the very best that can be achieved is an open problem that
involves solving the path-pruning problem globally. This problem is NP-hard as
the one solved by MRA, but its input, relating to the WMN in a global scale, is
much more sizable.

Another interesting aspect for further investigation is the effect of MRA on
multi-radio networks. In such schemes the network's path density can be
drastically reduced for a given frequency, thus potentially benefiting MRA.
Conversely, it may also be possible to use MRA to achieve the desirable goal of
minimizing the number of radios \cite{Bahl2004,Raniwala2005}. Further research
is also needed on the trade-off that clearly exists between obtaining spatially
separated paths or relatively short ones. While the former is good for
non-interference, disregarding the latter may lead to throughput loss on the
relatively longer paths.

\section*{Acknowledgments}

We acknowledge partial support from CNPq, CAPES, a FAPERJ BBP grant, and a
scholarship grant from Universit\'{e} Pierre et Marie Curie. All computational
experiments were carried out on the Grid'5000 experimental testbed, which is
being developed under the INRIA ALADDIN development action with support from
CNRS, RENATER, and several universities as well as other funding bodies (see
\texttt{https://www.grid5000.fr}).

\bibliography{mra}
\bibliographystyle{plain}

\newpage
\appendix
\setcounter{figure}{0}
\renewcommand\thefigure{\Alph{section}.\arabic{figure}}
\section{Supplementary figures}
\begin{figure}[t]
\centering
\scalebox{0.55}{\includegraphics{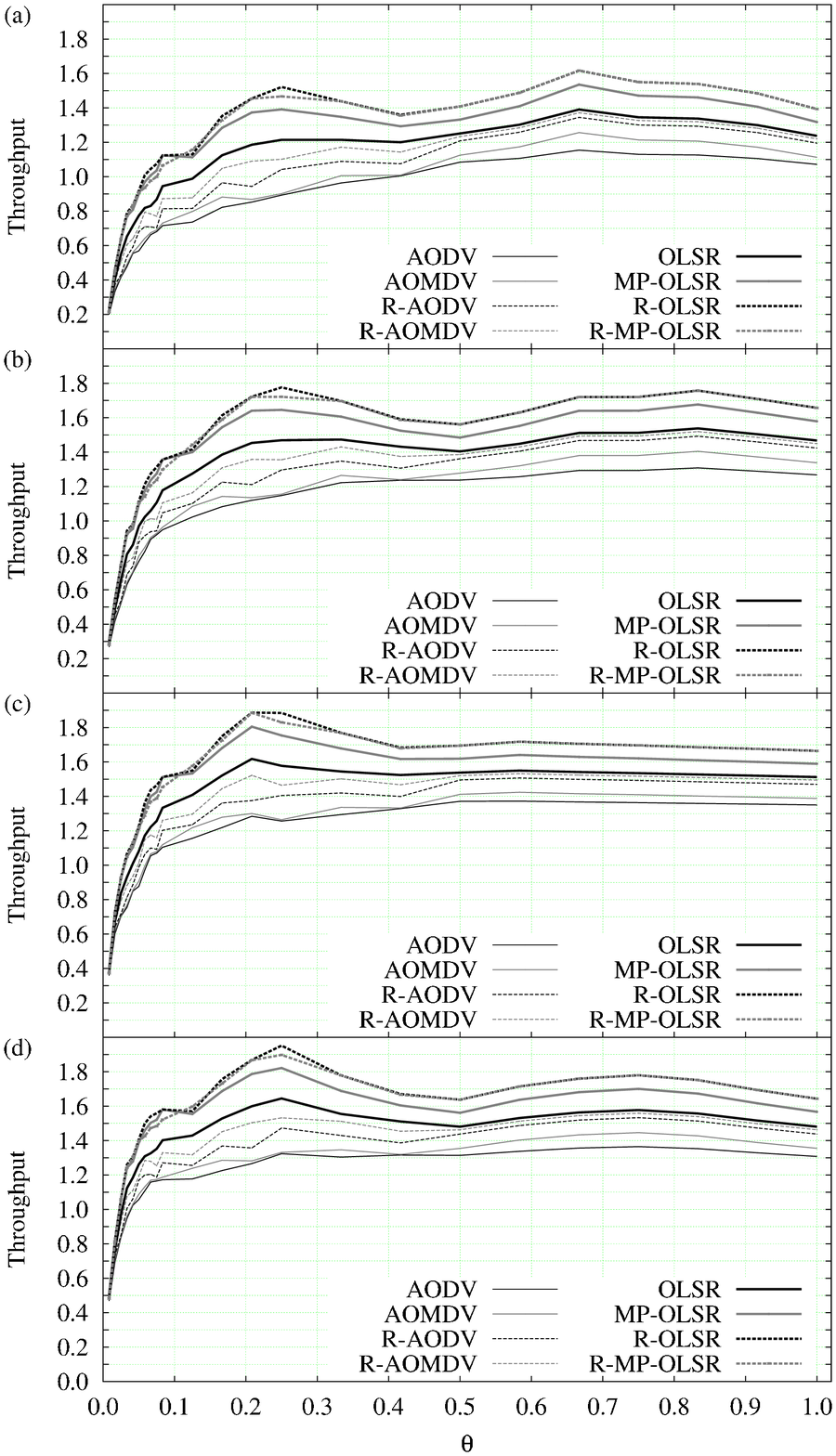}}
\caption[Throughput (packets/time slot) for CBR1]{Throughput (packets/time slot)
for CBR1 with $n=120$, for $\Delta=4$ (a), $\Delta=8$ (b), $\Delta=16$ (c), and
$\Delta=32$ (d). Data are averages over the $10^4$ path sets that correspond to
each value of $\Delta$ for each value of $\theta$. Confidence intervals are less
than $1\%$ of the mean at the $95\%$ level, so error bars are omitted.}
\label{figureS1}
\end{figure}

\begin{figure}[p]
\centering
\scalebox{0.55}{\includegraphics{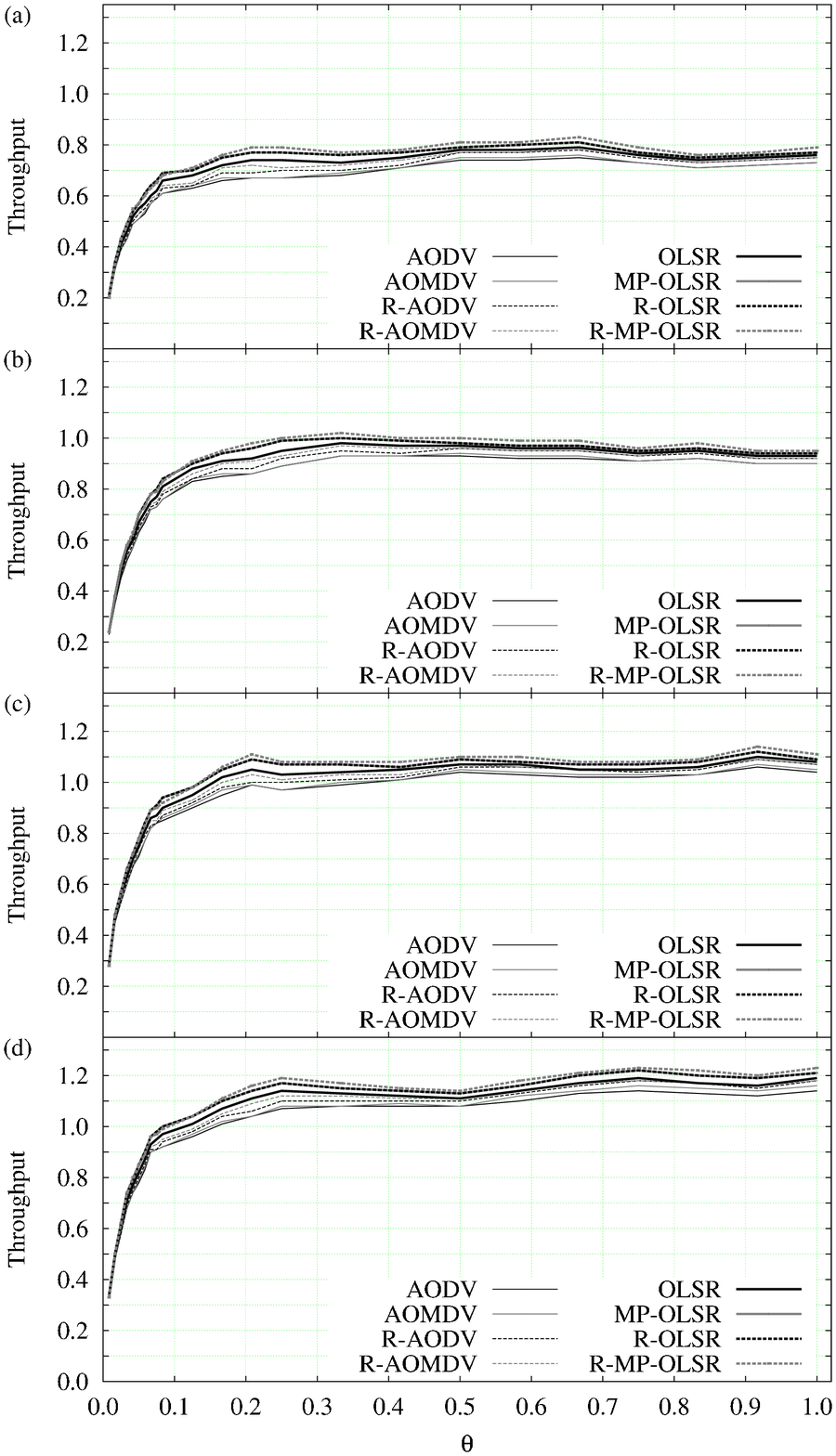}}
\caption[Throughput (packets/time slot) for CBR2]{Throughput (packets/time slot)
for CBR2 with $n=120$, for $\Delta=4$ (a), $\Delta=8$ (b), $\Delta=16$ (c), and
$\Delta=32$ (d). Data are averages over the $10^4$ path sets that correspond to
each value of $\Delta$ for each value of $\theta$. Confidence intervals are less
than $1\%$ of the mean at the $95\%$ level, so error bars are omitted.}
\label{figureS2}
\end{figure}

\begin{figure}[p]
\centering
\scalebox{0.55}{\includegraphics{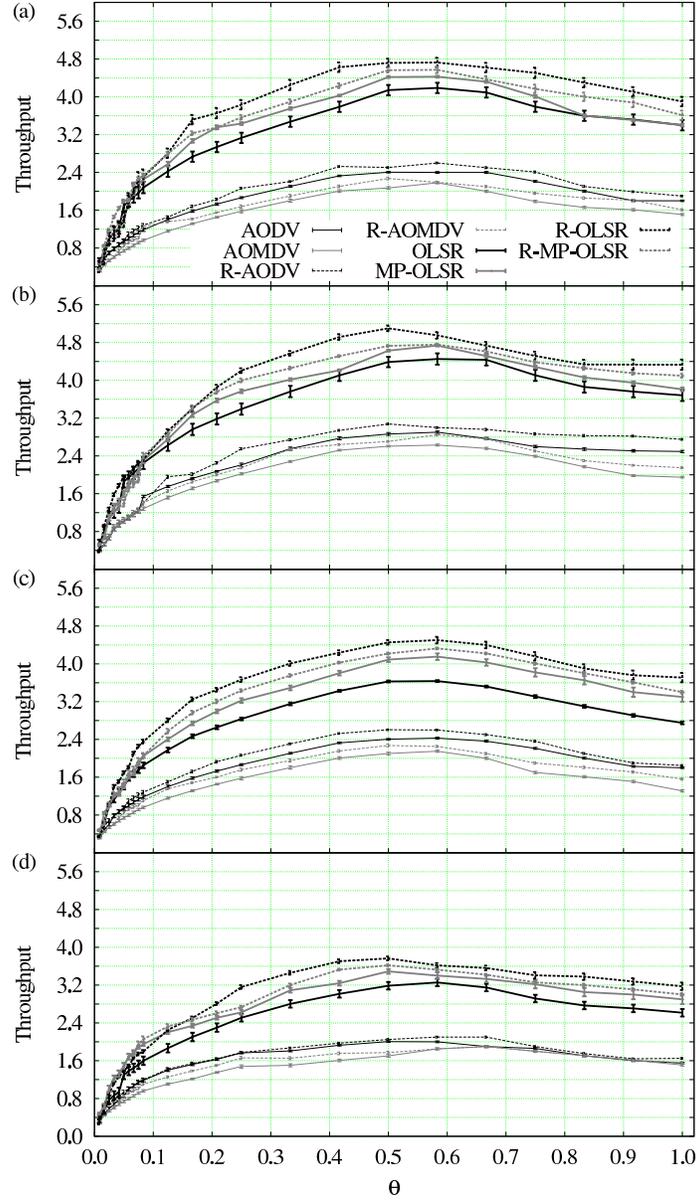}}
\caption[Throughput (packets/time slot) for SERA]{Throughput (packets/time slot)
for SERA with $n=120$, for $\Delta=4$ (a), $\Delta=8$ (b), $\Delta=16$ (c), and
$\Delta=32$ (d). Data are averages over the $10^4$ path sets that correspond to
each value of $\Delta$ for each value of $\theta$. Error bars are based on
confidence intervals at the $95\%$ level.}
\label{figureS3}
\end{figure}

\begin{figure}[p]
\centering
\scalebox{0.55}{\includegraphics{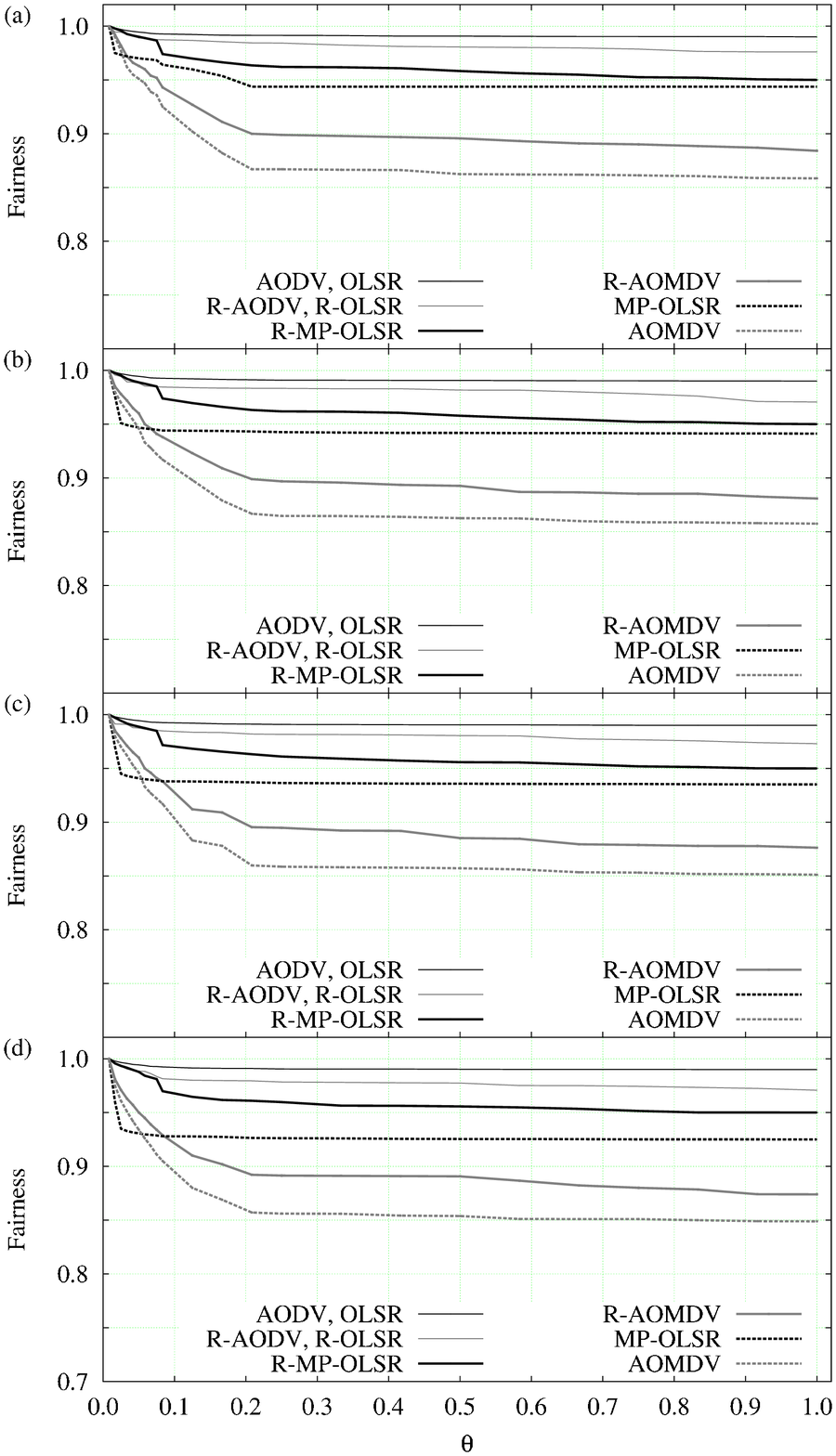}}
\caption[Alternative fairness index for SERA]{Alternative fairness index for
SERA with $n=120$, for $\Delta=4$ (a), $\Delta=8$ (b), $\Delta=16$ (c), and
$\Delta=32$ (d). Data are averages over the $10^4$ path sets that correspond to
each value of $\Delta$ for each value of $\theta$. Confidence intervals are
less than $1\%$ of the mean at the $95\%$ level, so error bars are omitted.}
\label{figureS4}
\end{figure}

\begin{figure}[p]
\centering
\scalebox{0.55}{\includegraphics{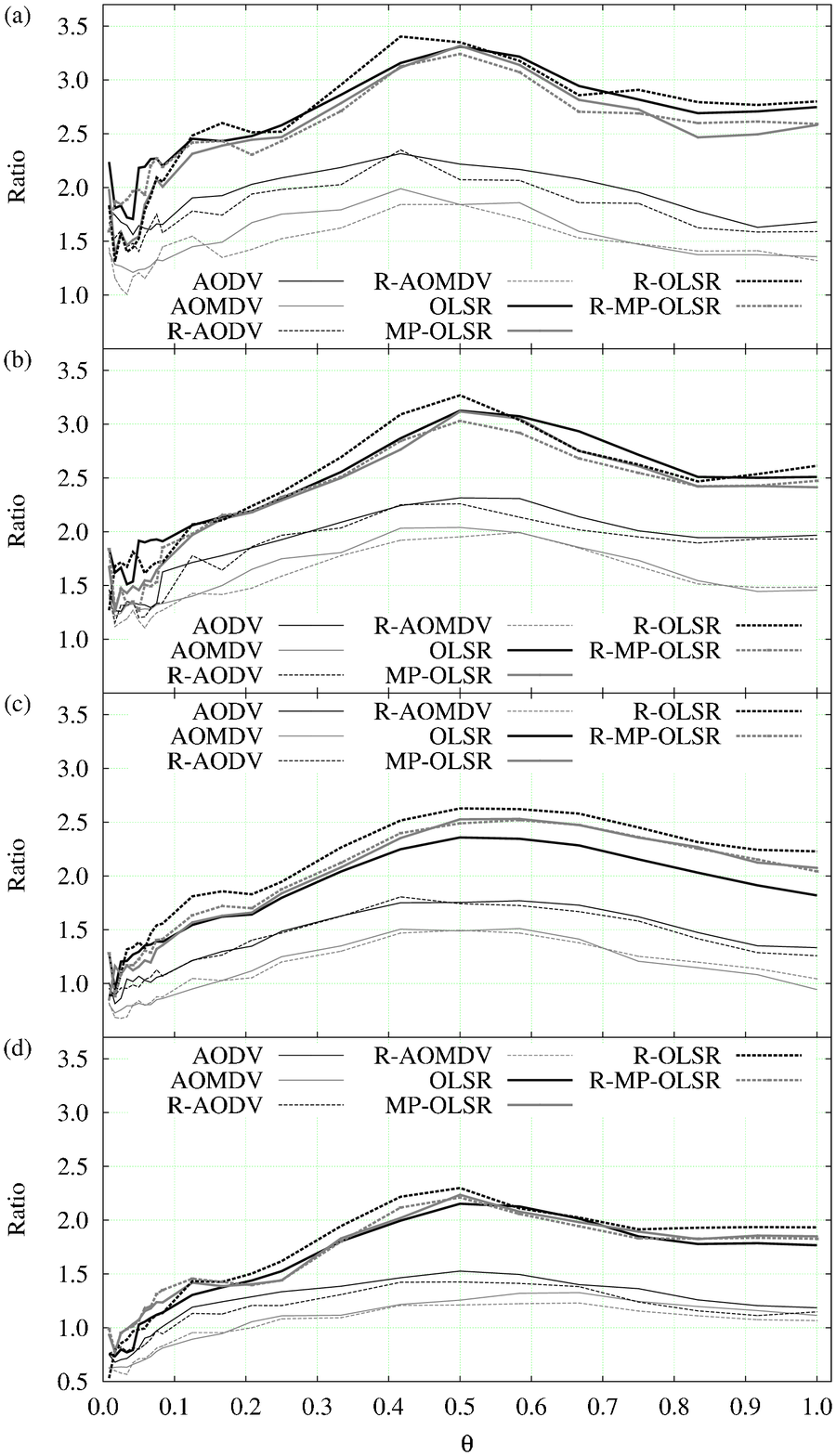}}
\caption[Ratio of SERA's throughput to that of CBR1]{Ratio of SERA's throughput
to that of CBR1 with $n=120$, for $\Delta=4$ (a), $\Delta=8$ (b), $\Delta=16$
(c), and $\Delta=32$ (d). Data are averages over the $10^4$ path sets that
correspond to each value of $\Delta$ for each value of $\theta$. Confidence
intervals are less than $1\%$ of the mean at the $95\%$ level, so error bars are
omitted.}
\label{figureS5}
\end{figure}

\begin{figure}[p]
\centering
\scalebox{0.55}{\includegraphics{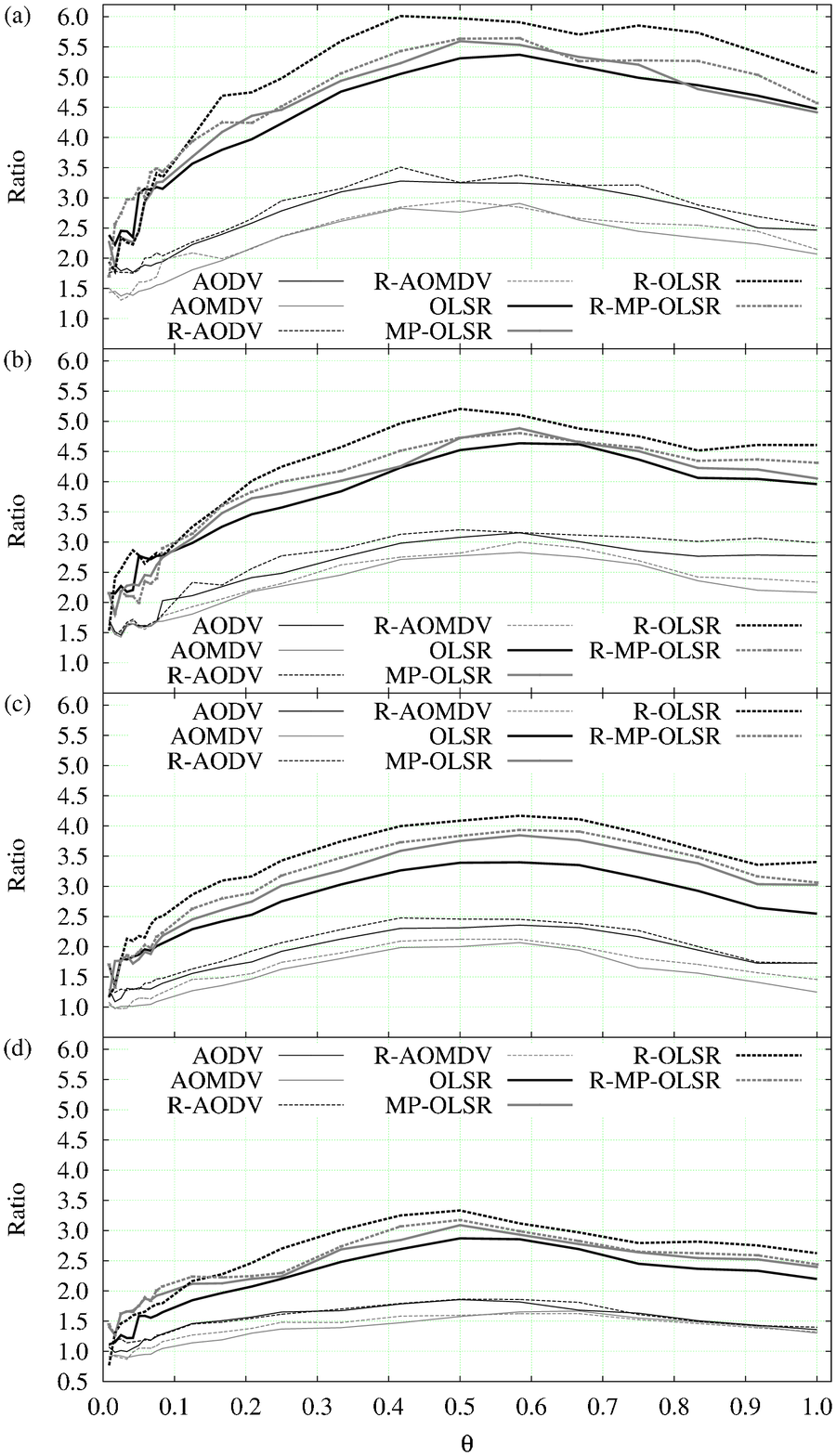}}
\caption[Ratio of SERA's throughput to that of CBR2]{Ratio of SERA's throughput
to that of CBR2 with $n=120$, for $\Delta=4$ (a), $\Delta=8$ (b), $\Delta=16$
(c), and $\Delta=32$ (d). Data are averages over the $10^4$ path sets that
correspond to each value of $\Delta$ for each value of $\theta$. Confidence
intervals are less than $1\%$ of the mean at the $95\%$ level, so error bars are
omitted.}
\label{figureS6}
\end{figure}

\begin{figure}[p]
\centering
\scalebox{0.55}{\includegraphics{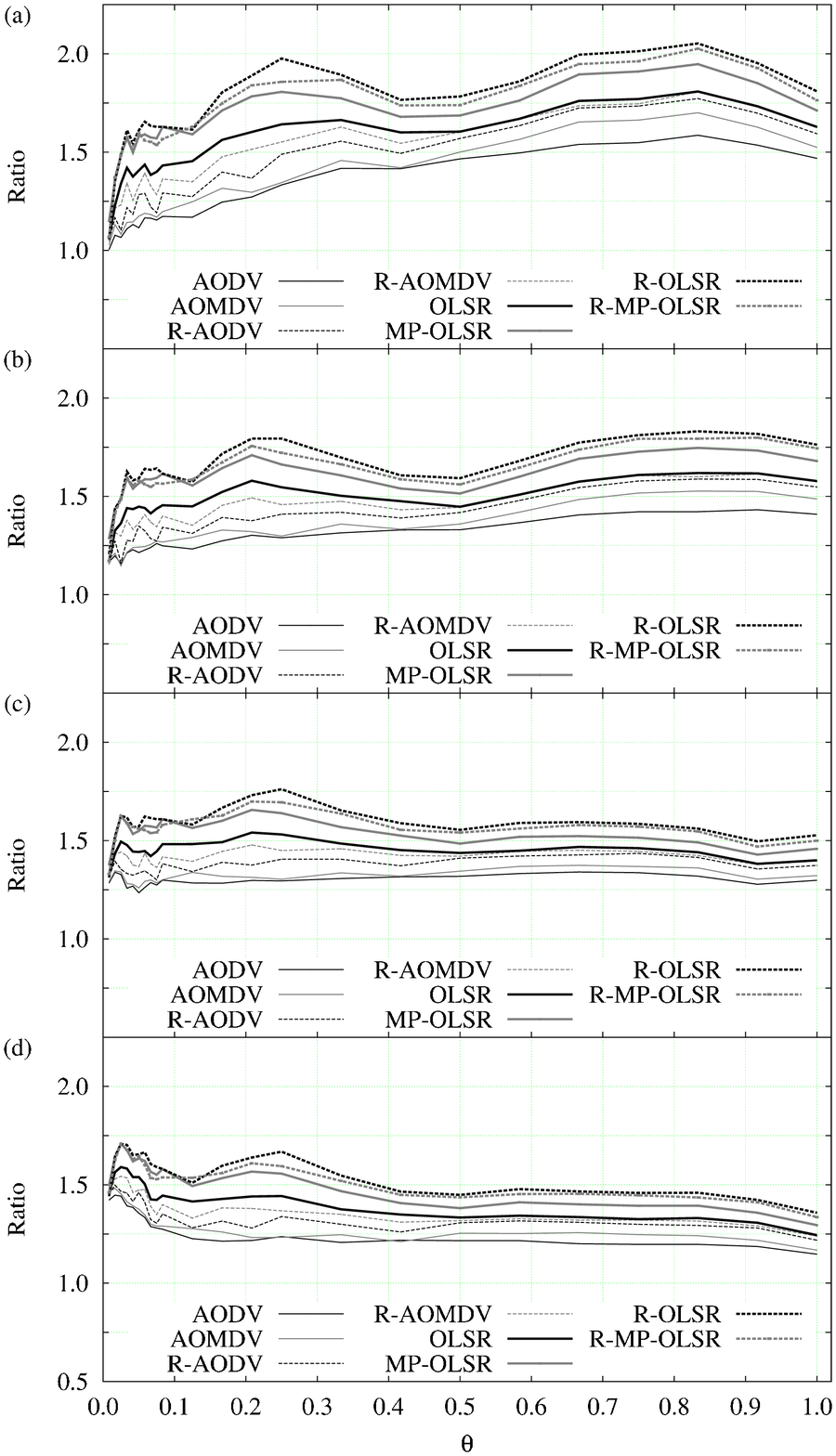}}
\caption[Ratio of CBR1's throughput to that of CBR2]{Ratio of CBR1's throughput
to that of CBR2 with $n=120$, for $\Delta=4$ (a), $\Delta=8$ (b), $\Delta=16$
(c), and $\Delta=32$ (d). Data are averages over the $10^4$ path sets that
correspond to each value of $\Delta$ for each value of $\theta$. Confidence
intervals are less than $1\%$ of the mean at the $95\%$ level, so error bars are
omitted.}
\label{figureS7}
\end{figure}

\begin{figure}[t]
\centering
\scalebox{0.55}{\includegraphics{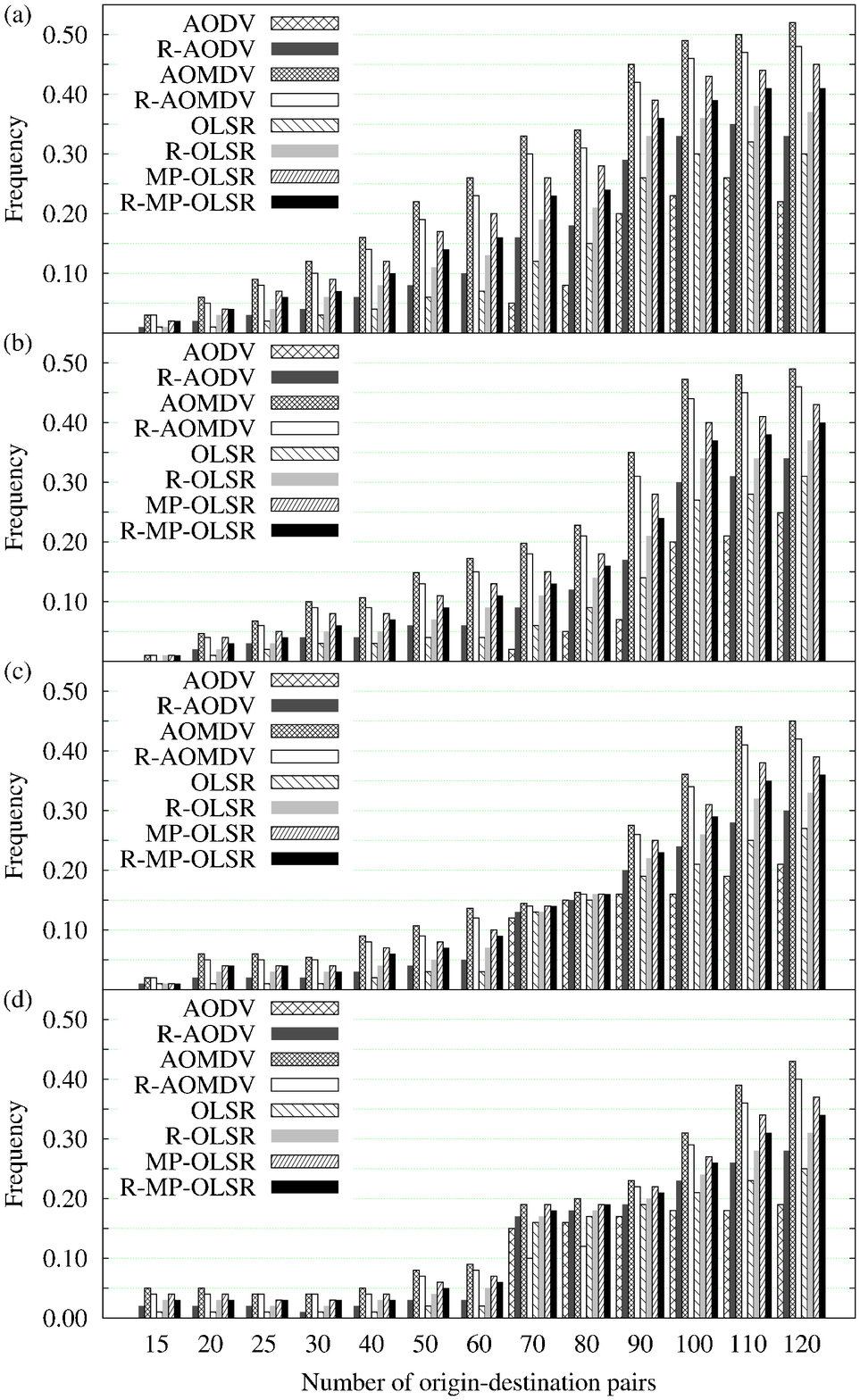}}
\caption[Origin-destination pairs without any traffic in CBR1]{Distribution of
the number of origin-destination pairs without any traffic in CBR1 with $n=120$,
for $\Delta=4$ (a), $\Delta=8$ (b), $\Delta=16$ (c), and $\Delta=32$ (d). Data
are averages over the $10^4$ path sets that correspond to each value of
$\Delta$.}
\label{figureS8}
\end{figure}

\begin{figure}[t]
\centering
\scalebox{0.55}{\includegraphics{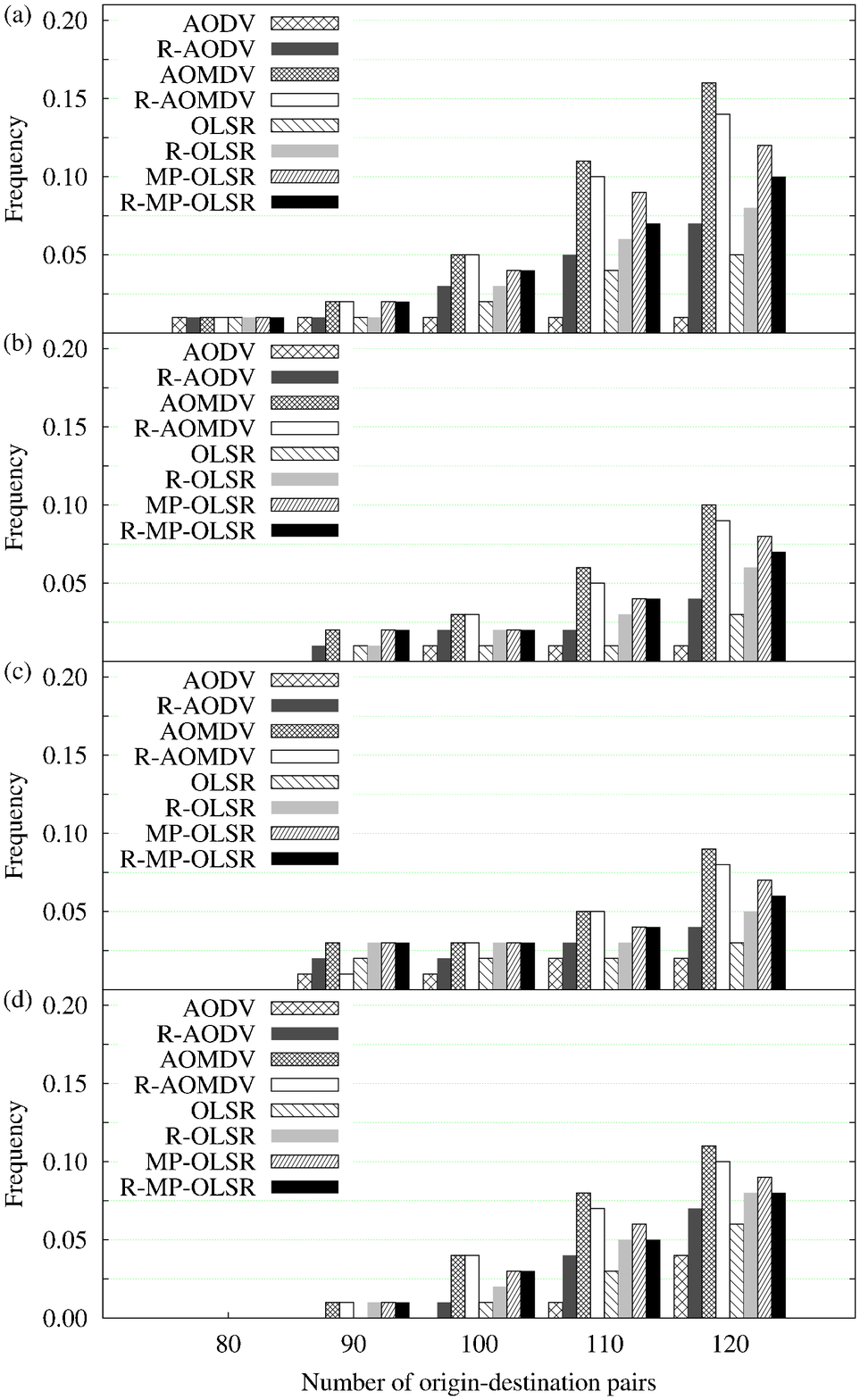}}
\caption[Origin-destination pairs without any traffic in CBR2]{Distribution of
the number of origin-destination pairs without any traffic in CBR2 with $n=120$,
for $\Delta=4$ (a), $\Delta=8$ (b), $\Delta=16$ (c), and $\Delta=32$ (d). Data
are averages over the $10^4$ path sets that correspond to each value of
$\Delta$.}
\label{figureS9}
\end{figure}

\end{document}